\documentclass[pre]{revtex4}
\usepackage{amsmath}
\usepackage[dvips]{graphicx}

\begin{document}

\preprint{APS/123-QED}

\title{Non-uniform state space reconstruction and coupling detection}

\author{Ioannis Vlachos}
\email{ivlaxos@gen.auth.gr}
\author{Dimitris Kugiumtzis}%
 \email{dkugiu@gen.auth.gr}
 \affiliation{Department of Mathematical, Physical
and Computational Sciences,\\ Faculty of Engineering Aristotle University of
Thessaloniki}

\date{\today}

\begin{abstract}
We investigate the state space reconstruction from multiple time series derived from
continuous and discrete systems and propose a method for building embedding vectors
progressively using information measure criteria regarding past, current and future
states. The embedding scheme can be adapted for different purposes, such as mixed
modelling, cross-prediction and Granger causality. In particular we apply this method
in order to detect and evaluate information transfer in coupled systems. As a
practical application, we investigate in records of scalp
epileptic EEG the information flow across brain areas.
\end{abstract}

\pacs{05.45.Tp, 02.50.Sk, 05.45.Xt, 87.19.lo}
\maketitle

\section{Introduction}\label{ivlach:sec1}

Since its publication, Takens' embedding theorem \cite{Takens81,Sauer91} is reputed as
the most significant tool in non-linear time series analysis and has been used in many
different settings ranging from system characterization and approximation of invariant
quantities to prediction and noise-filtering \cite{Kantz97}. The embedding theorem
implies that although the true dynamics of a system may not be known, equivalent dynamics
can be obtained using time delays of a single time series, seen as the one-dimensional
projection of the system trajectory.

Most applications of the embedding theorem deal with univariate time series, but often measurements of more than one quantities related to the same
dynamical system are available. One of the first uses of multivariate embedding was in the context of spatially distributed systems, where embedding
vectors were formed from simultaneous measurements of a variable at different locations \cite{Guckenheimer83,Palus92}. Multivariate embedding was
used among others for noise reduction \cite{Hegger92} and in the the test for nonlinearity using multivariate surrogate data \cite{Prichard94}.
Another particular implementation was the prediction of a time series using local models on a state space reconstructed from a different time series
of the same system \cite{Abarbanel94} and with prediction criteria for the determination of the embedding vectors. In a later work, embedding was
extended to all of the observed time series \cite{Cao98} with the goal again being the prediction of one time series. Finally, multivariate embedding
has been used widely for the estimation of coupling strength and direction in coupled systems \cite{Feldmann2004,Romano2007,Faes08,Janjarasjitt2008},
where the embedding parameters were chosen either arbitrarily or separately for each time series.

Regarding state space reconstruction, in \cite{barnard01} independent components analysis
was used to extrapolate from an embedding space created from multiple variables a
projection with linear independent variables, bypassing thus the problem of optimal
parameters selection. In \cite{boccaletti02} a modified false nearest neighbors approach
was used for creating the embedding vectors and was used to disentangle subspace
dimensionalities in weakly coupled dynamical systems. Recently, non-uniform multivariate
embedding using different delays was addressed in \cite{garcia05,pecora07}, with ideas
based again on false nearest neighbors.

Most of these works are basically extensions of methods used in the scalar case
without taking into account particular characteristics of
multivariate embedding, such as the possible high dependence of individual components
obtained from the different time series. As in the univariate case, the optimal embedding
vectors depend highly on the purpose of the reconstruction.

In this work, we contemplate on the particularities of multivariate embedding and suggest
a new embedding scheme to treat them. Based on Fraser's idea of using information
criteria to select the embedding parameters in the univariate case
\cite{fraser891,fraser892}, we propose a method for progressive building of the embedding
vectors allowing for different variables and delays that is simple, intuitively sound and
can easily be modified to handle most purposes of univariate and multivariate analysis.
For bivariate time series, we also propose two causality measures based on the mixed
embedding scheme.

In Sec.~\ref{ivlach:sec2}, the state space reconstruction from univariate and
multivariate time series is briefly discussed, and in Sec.~\ref{ivlach:sec3}, the
proposed embedding scheme is presented together with  a new method for the estimation
of coupling strength and direction. In Sec.~\ref{ivlach:sec4} results from simulations on
well known coupled systems are given, and in Sec.~\ref{ivlach:sec5} the proposed scheme
is applied to investigate information flow in the brain of epileptic patients.
Finally, summary results are discussed in Sec.~\ref{ivlach:sec6}.

\section{State space reconstruction}
\label{ivlach:sec2}

We suppose that a trajectory is generated by a dynamical system forming an attractor and
we observe projections of the trajectory in one or more dimensions, i.e. we have
measurements of a univariate or multivariate time series. The objective is to form a
pseudo-state space in a way that as much information as possible about the original dynamics
can be represented in the reconstructed points in this space.

\subsection{Univariate time series}

For a given time series $\{x_n\}_{n=1}^N$, the uniform embedding scheme is directly
derived from Takens' embedding theorem \cite{Takens81} as
\begin{equation}
\label{ivlach:embvec} \mathbf{x}_n=(x_n,x_{n-\tau},...,x_{n-(m-1)\tau}).
\end{equation}
The two parameters of the delay embedding in \eqref{ivlach:embvec} are the embedding
dimension $m$, i.e. the number of components in $\mathbf{x}_n$, and the delay time $\tau$.
According to Takens' theorem topological equivalence of the original and the
reconstructed attractor can be established (under generic conditions) if $m \ge 2d+1$,
where $d$ is the fractal dimension of the original attractor. Among the approaches for
the selection of $m$ the most popular are the \emph{false nearest neighbors} (FNN)
\cite{Kennel92} and the goodness-of-fit or prediction criteria \cite{Kaplan93,Chun-Hua04},
while $\tau$ is usually chosen as the one that gives the first local minimum of the
time-delayed mutual information \cite{Fraser86} or the first zero of the autocorrelation
function \cite{Albano87}.

Judd and Mees \cite{Judd98} introduced the non-uniform embedding scheme for univariate
time series using unequal lags, so that
$\mathbf{x}_n=(x_{n-l_1},x_{n-l_2},\ldots,x_{n-l_m}),$
and used it to obtain global reduced autoregressive models with the help of basis functions.
The embedding vectors were constructed progressively by a \emph{minimum description length}
(MDL) criterion \cite{Rissanen78}. In \cite{Small03} again MDL and model prediction error, and in
\cite{garcia05b} a false nearest neighbors statistic, were used to construct the embedding vectors.

\subsection{Multivariate time series}

Given that there are $p$ time series $\{x_{i,n}\}_{n=1}^N$, $i=1,\ldots,p$, generated
through $p$ different projections of the trajectory of the original dynamical system, the
simplest extension of the uniform reconstructed state space vector in
\eqref{ivlach:embvec} is of the form
\begin{equation}\label{eq:simplemulembvec}
\mathbf{x}_n=(x_{1,n},x_{1,n-\tau},\ldots,x_{1,n-(m_1-1)\tau},x_{2,n},\ldots,x_{p,n-(m_p-1)\tau})
\end{equation}
and is defined by an \emph{embedding dimension vector} $\mathbf{m}=(m_{1},...,m_{p})$
that indicates the number of components used from each time series. The dimension of the
reconstructed space in this case is $M=\sum\limits_{i=1}^p m_i$. Also one can choose to
use different $\tau$ for each time series, defining so a \emph{time delay vector}
${\pmb\tau}=(\tau_{1},...,\tau_{p})$.

\subsection{Problems and restrictions of multivariate state space reconstruction}

A major problem in the multivariate case is the \emph{problem of identification}. There is often not a unique $\mathbf{m}$ that unfolds fully the
attractor, even for a fixed $M$. For the univariate case, given a value for $\tau$ and the minimum embedding dimension $m=2d+1$, the embedding
vectors are uniquely defined. Here, theoretically, all different combinations of $m_i$ that fulfill $M= 2d+1$ would fully unfold the attractor
\cite{Sauer91}, provided that all the variables are from the same system. In practice things are not so simple. Suppose we have a method for
comparing these reconstructions and selecting the optimal one. Due to finite data size and resolution one of these reconstructions would be deemed
marginally better than the others. If we add small observational noise to the data and compare again the reconstructions, the one selected as optimum
would most likely change. We elaborate more on this when we discuss the proposed embedding scheme below.

An issue of concern is also the fact that multivariate data do not always have the same range, so that distances calculated on delay vectors may
depend highly on the components of large ranges. So, it is often preferred to scale all the data to have either the same variance or be at the same
data range \cite{Kantz97}. In our study, we choose to scale the data to the range $[0,1]$.

A different problem of multivariate reconstruction is that of {\em irrelevance}, when
time series that are not generated by the same dynamical system are included in the
reconstruction procedure. This may be the case even when for two time series connected to
each other through the observation function, only one of them is generated by the system
under investigation.

The opposite problem to irrelevance is that of {\em redundancy}. Often the observed variables
may be synchronous or lagged correlated and this may add overlayed information in the embedding
vector representation.
In linear modeling, the problem of redundancy has been addressed using appropriate order selection
and regularization techniques \cite{Jolliffe02,Burnham02}, and these tools have been used in
local state space models for univariate time series prediction \cite{Sauer94,Kugiumtzis98}.
Nonlinear redundancy measures have been existing for long time
(e.g. \cite{fraser891,Palus93c} and references therein) and have been used in univariate
time series, e.g. for the selection of optimal delay $\tau$.
Still, when selecting $\tau$, so that two consecutive components of the embedding vector
are not correlated to each other, this does not establish that non-consecutive components are
also not correlated \cite{Kugiumtzis96}.
Thus the problem of redundant information should be treated collectively for all
components of the embedding vector and not only for pairs of components,
which constitutes a difficult task.
The situation is even more difficult for multivariate time series, where redundancy should
be investigated for both different and lagged variables. Our approach is aiming at treating this
problem.

\section{Proposed state space reconstruction}\label{ivlach:sec3}

\subsection{Non-uniform multivariate reconstruction}

Usage of the non-uniform scheme for multivariate embedding may reduce better the redundant information than the embedding with fixed lags. The most
general embedding scheme allowing for $m_i$ varying lags $l_{ij}$, $j=1,\ldots,m_i$ at each component variable $x_i$ is
\[
 \mathbf{x}_n=(x_{1,n-l_{11}},x_{1,n-l_{12}},\ldots,x_{1,n-l_{1m_1}},x_{2,n-{l_{21}}},\ldots,x_{p,n-l_{pm_p}}).
\]

Inspection of all possible combinations of components for the determination of the optimum
embedding (with any criterion we may choose) can be computationally intractable when the
embedding dimension and the number of time series are large. For a moderate example when
$p$ = 3, $M$ = 3, and the maximum lag for all time series is 10 we have 4050 cases, while
when we increase $M$ to 4 we have 27405 cases. It is evident that for most practical
purposes we need a progressive method to build the embedding vectors.

Assume that we have somehow chosen the maximum lags $L_{i}$ for each time series
$i=1,\ldots,p$. Then we have a collective set of the $L_1 + L_2 + \ldots + L_p$ candidate
components
$$\mathbf{B}=\{x_{1,n},x_{1,n-1},\ldots,x_{1,n-L_{1}},x_{2,n},\ldots,x_{p,n-L_{p}}\}$$
and the embedding vector $\mathbf{b}$ to be selected is a subset of $\mathbf{B}$. Using
ideas developed for the state space reconstruction from univariate time series, the
reconstructed vector must satisfy two properties. First, its components must be least
dependent to each other, and secondly, it must be able to explain best the dynamics of
the system, namely its future (or better our knowledge about its future) for one, two or
several steps ahead. At time $n$ the future of the system is represented by the vector
$$\mathbf{x}_{F}=(x_{1,n+1},x_{1,n+2},\ldots,x_{1,n+T_1},x_{2,n+1},\ldots,x_{p,n+T_p}),$$
for appropriate time horizon $T_i$ for each variable $x_i, i=1,\ldots,p$.

The proposed scheme is progressive, and starting from an empty vector $\mathbf{b}_0$,
suppose that at step $j$ we have already selected the $j-1$ components $\mathbf{b}_{j-1}
=(x^{\star}_1,x^{\star}_2,\ldots,x^{\star}_{j-1})$. Then we add the component $x^{\star}_j\in
\mathbf{B}\setminus\{x^{\star}_1,\ldots,x^{\star}_{j-1}\}$ that fulfills the following
maximization criterion
\begin{equation}
\label{ch6:simplecriterion}
\max\limits_{x_j}\left\{I\big(\mathbf{x}_{F};\big(x_j,\mathbf{b}_{j-1}\big)\big)\right\},
\end{equation}
where $I\big(\mathbf{x}_{F};\big(x_j,\mathbf{b}_{j-1}\big)\big)$ is the mutual
information between the variable $\mathbf{x}_{F}$ that represents the future of the system and the new
embedding vector $\big(x_j,\mathbf{b}_{j-1}\big)$. The maximum value
$I\big(\mathbf{x}_{F};\big(x^{\star}_j,\mathbf{b}_{j-1}\big)\big)$ regards the largest
amount of information on $\mathbf{x}_{F}$ that can be gained by the augmented embedding
vector, and thus $\mathbf{b}_{j}=\big(x^{\star}_j,\mathbf{b}_{j-1}\big)$.

As a stopping criterion for this scheme of progressive vector building we use a threshold
for the change in the mutual information of $\mathbf{x}_{F}$ and the embedding vectors in
the two last steps. We stop at step $j$ and use $\mathbf{b}_{j-1}$ as optimal if
\begin{equation}\label{ch6:threshold}
I\big(\mathbf{x}_{F};\mathbf{b}_{j-1}\big)/I\big(\mathbf{x}_{F};\mathbf{b}_{j}\big)> A,
\end{equation}
where $A$ is a threshold value, and $A \le 1$. A threshold near 1, e.g. $A=0.95$, allows
the inclusion of a new component in the embedding vector even if the augmented vector
explains very little of the information on $\mathbf{x}_{F}$ that was not explained at the
previous step.

Depending on the form of $\mathbf{B}$ and $\mathbf{x}_F$ this approach can be used for
univariate modelling when both $\mathbf{B}$ and $\mathbf{x}_F$ contain elements from the
same time series, cross modelling when $\mathbf{x}_F$ has elements from one time series
and $\mathbf{B}$ from another, mixed modelling if $\mathbf{x}_F$ has elements from one
time series and $\mathbf{B}$ from all time series and ``full'' modelling if
$\mathbf{x}_F$ and $\mathbf{B}$ have elements from all time series. In this way, we give
a purpose to the embedding vectors that we create. If one wants to check relations
between the time series, cross modelling and mixed modelling should be preferred, while
for the estimation of invariant quantities of the underlying system a full modelling
would be better.

The mutual information regards vector variables and thus the estimation is more
difficult than for the standard delayed mutual information used to estimate the fixed lag
$\tau$ for univariate time series. Thus binning estimates of mutual information are not
appropriate here as data demand increases dramatically with the dimension, at the level
of millions of data points for large dimension. Since our embedding scheme depends
heavily on the accurate estimation of information measures, the estimation method is of
outmost importance.

\subsection{Mutual information estimation}

In \cite{Kraskov04}, the mutual information (MI) of two variables $X$ and $Y$ of dimensions
$d_X$ and $d_Y$, respectively, was estimated from a sample of length $N$ based on the
well known formula
\begin{eqnarray}\label{ivlach:e1}
I(X,Y)=H(X)+H(Y)-H(X,Y),
\end{eqnarray}
with entropies $H$ estimated from $K$-nearest neighbor distances (for details, see \cite{Kraskov04}).

First the joint entropy was estimated by
\begin{eqnarray}\label{ivlach:e2}
\hat H(X,Y)=-\psi(k)+\psi(N)+\log(c_{d_X}c_{d_Y})+{d_X+d_Y \over
N}\sum\limits_{i=1}^{N}\log\mathrm{\epsilon}(i),
\end{eqnarray}
where $\mathrm{\epsilon}(i)$ is twice the distance of the $i$-th sample point in the
joint space $(X,Y)$ to its $K$-th neighbor, $\psi(x)$ is the digamma function
($\psi(x)={d \over dx}\log{\Gamma(x)}={\Gamma ' (x) \over \Gamma (x)}$) and $c_{d}$ is
the volume of the $d$-dimensional unit cube, using maximum norm as the distance metric. In order to
obtain the entropy $H(X)$ the space of variable $X$ is treated as a projection from the
joint space and so
\begin{eqnarray}\label{ivlach:e3}
\hat H(X)=-{1 \over
N}\sum\limits_{i=1}^{N}\psi[n_{x}(i)]+\psi(N)+\log(c_{d_X})+{d_X \over
N}\sum\limits_{i=1}^{N}\log\mathrm{\epsilon}(i),
\end{eqnarray}
with $n_{x}(i)$ the number of points whose distance from the $i$-th point of $X$ is less
than $\mathrm{\epsilon}(i)/2$, plus one. From Eq.~\eqref{ivlach:e1}-\eqref{ivlach:e3} and
denoting by $\big<\ldots\big>$ the average over all $i$ immediately follows
\begin{eqnarray}\label{ivlach:e4}
\hat I(X;Y)=\psi(k)+\psi(N)-\big<\psi[n_{x}(i)]+\psi[n_{y}(i)]\big>.
\end{eqnarray}

Similarly to other estimates of MI, this one suffers from bias that
depends heavily on the dimension of the vector variables. The setup of the embedding
procedure allows us to use the \emph{conditional mutual information} (CMI) instead that, as
will be explained further below, gives smaller bias. Assume three
time series $X$, $Y$ and $Z$ of length $N$ with dimension $d_X$, $d_Y$ and $d_Z$.
CMI is defined as
\begin{eqnarray}\label{ivlach:e5}
I(X;Y \mid Z)&=&I(X;(Y,Z))- I(X;Z)= \\\nonumber
 &=&H(X,Z)+H(Y,Z)-H(Z)-H(X,Y,Z).
\end{eqnarray}

Following the above method, we estimate the joint entropy $H(X,Y,Z)$ using $K$-nearest
neighbors and obtaining a formula similar to Eq.~\eqref{ivlach:e2}. Then the entropies
$H(X,Z)$, $H(Y,Z)$ and $H(Z)$ are estimated by projecting the space of ($X,Y,Z$) onto the respective
subspaces of ($X,Z$), ($Y,Z$) and $Z$, obtaining formulas similar to
Eq.~\eqref{ivlach:e3}. After substitution, we get the $K$-nearest neighbor estimate for
CMI as
\begin{eqnarray}\label{CMIkraskov}
\hat I(X;Y \mid Z)&=&\psi(k)-\big<\psi[n_{xz}(i)]+\psi[n_{yz}(i)]-\psi[n_{z}(i)]\big>,
\end{eqnarray}
where $n_{xz}(i)$ is the number of points whose distance from the $i$-th point of ($X,Z$) is less than $\mathrm{\epsilon}(i)/2$, plus one, and the same
for $n_{yz}(i)$.

An alternative derivation of the expression of $\hat I(X;Y \mid Z)$ is from the difference
of the estimates of the two MI in Eq.~\eqref{ivlach:e5}. The first MI is estimated
from Eq.~\eqref{ivlach:e4} for the variables $X$ and $(Y,Z)$ as
\begin{eqnarray}\label{ivlach:e7}
\hat I(X;(Y,Z))=\psi(k)+\psi(N)-\big<\psi[n_{x}(i)]+\psi[n_{yz}(i)]\big>.
\end{eqnarray}
Note that the second MI is not estimated independently, i.e. from Eq.~\eqref{ivlach:e4}
for the variables $X$ and $Z$, but through the projection of the space of $(X,Y,Z)$ onto
the subspace of $(X,Z)$. Using the definition of MI in Eq.~\eqref{ivlach:e1} and the
estimate of the projected entropy in Eq.~\eqref{ivlach:e3}, we get
\begin{eqnarray}\label{ivlach:e8}
\hat I_{p}(X;Z)=\psi(N)-\big<\psi[n_{x}(i)]+\psi[n_{z}(i)]-\psi[n_{xz}(i)]\big>.
\end{eqnarray}

Our embedding scheme relies on the estimates in Eq.~\eqref{CMIkraskov}, Eq.~\eqref{ivlach:e7}
and Eq.~\eqref{ivlach:e8}. Therefore we will examine the bias of these estimates.
Possible factors for the bias of the MI estimate in Eq.~\eqref{ivlach:e7}
and Eq.~\eqref{ivlach:e8} are the number of available data $N$, the
correlation structure for the variables denoted loosely as $c$ and the dimension of the joint
variable space $d$ \cite{Kraskov04}. In our approach, $d_Y=1$ and the estimate $\hat I(X;(Y,Z))$ of the
real $I(X;(Y,Z))$ would be
\begin{equation}
\hat I(X;(Y,Z)) = I(X;(Y,Z)) + B(N,d_x+d_z+1,c)\label{eqMI1},
\end{equation}
where $B(N,d_x+d_z+1,c)$ is the unknown bias term. Assuming that $\hat I(X;Y \mid
Z)=\hat I(X;(Y,Z))- \hat I(X;Z)$ and $B_1$, $B_2$ are the bias of the first and second
term, the estimate for CMI is
\begin{eqnarray}\nonumber
\hat I(X;Y \mid Z) &=& I(X;(Y,Z)) + B_1(N,d_x+d_z+1,c)\\\nonumber &&-I(X;Z) - B_2(N,d_x+d_z,c')\\\nonumber
                    &=& I(X;Y \mid Z)+  B_1(N,d_x+d_z+1,c)\\
                    && - B_2(N,d_x+d_z,c')\label{eqMI2}
\end{eqnarray}
Since in Eq.~\eqref{eqMI2} we have the difference of two bias terms we expect that if MI
is systematically underestimated or overestimated, the CMI estimate will have reduced
bias.

The CMI estimate in Eq.~\eqref{CMIkraskov} truly has reduced bias, as a limited
simulation study revealed. We tested three indicative cases for Gaussian multivariate
time series of zero mean, unit variance and varying dimension and correlation structure. The
multivariate time series regard the variable set ($X,Z,Y$) of dimensions ($d_X,d_Z,d_Y$)
respectively, and in each of the three cases we fix the dimensions of $X$ and
$Y$, and vary the dimension of $Z$ and the correlation structure between the
variables. With reference to the embedding procedure, $X$ corresponds to $\mathbf{x}_{F}$,
$Z$ to $\mathbf{b}_{j-1}$ and $Y$ to $x_j$ ($d_Y=1$ for all cases studied in accordance
with the embedding procedure).

For the first case, we choose the ($X,Z,Y$) time series from a 10-dimensional time series $W$ with
correlation matrix
\begin{equation}
\nonumber R= \left[ \begin{array}{cccc}
1       & r         & \cdots &  r \\
r \ & 1   & \cdots & r \\
\vdots  & \vdots    & \ddots & \vdots\\
r   & r     & \cdots & 1
\end{array} \right]
\end{equation}
as follows. For $d_Z=1$, we take the first 3 time series with correlation matrix given
by the top--right $3 \times 3$ block of $R$, denote the first one as $X$, the
second as $Z$ and the last as $Y$. For $d_Z=2$ we take the first 4 time series, denote
the first again as $X$, the 2 next as the 2-dimensional $Z$ and the last as $Y$ and so on
for $d_Z=3,\ldots,8$. The cross-correlation is the same among all variables at a level $r$
ranging from 0.2 to 0.9 with step 0.1. With this setup we want to study the effect of the
change in dimension of $Z$ ($\mathbf{b}_{j-1}$)
to the MI and CMI estimation when the time series are correlated to each other at the
same degree.

In the second case, $W$ has dimension 17 and $R$ is as above but with size $17\times 17$.
Thus the only change is that $X$ is 8-dimensional comprised of the first 8 time series of $W$, while
$d_Y=1$, $d_Z$ ranges again from 3 to 8, and $r$ from 0.2 to 0.9. In this way we study the
same effect as before in the presence of a larger dimension for variable $X$ (larger
dimension for $\mathbf{x}_{F}$).

For the third case, $W$ is 10-dimensional with correlation matrix
\begin{equation}
\nonumber R= \left[ \begin{array}{cccc}
1       & 0.9         & \cdots &  0.2 \\
0.9 \ & 1   & \cdots & 0.3 \\
\vdots  & \vdots    & \ddots & \vdots\\
0.2   & 0.3     & \cdots & 1
\end{array} \right]
\end{equation}
and the setup is as in the first case. Here we want to study the dimension effect when the
correlation between the time series is varying.  This simulation is more in context
(although by no means identical) to the progressive procedure, where we would expect
first the most relevant lag to be selected, then the second more relevant etc.

\begin{figure}[htb]
\centerline{\hbox{\includegraphics[height=3.3cm]{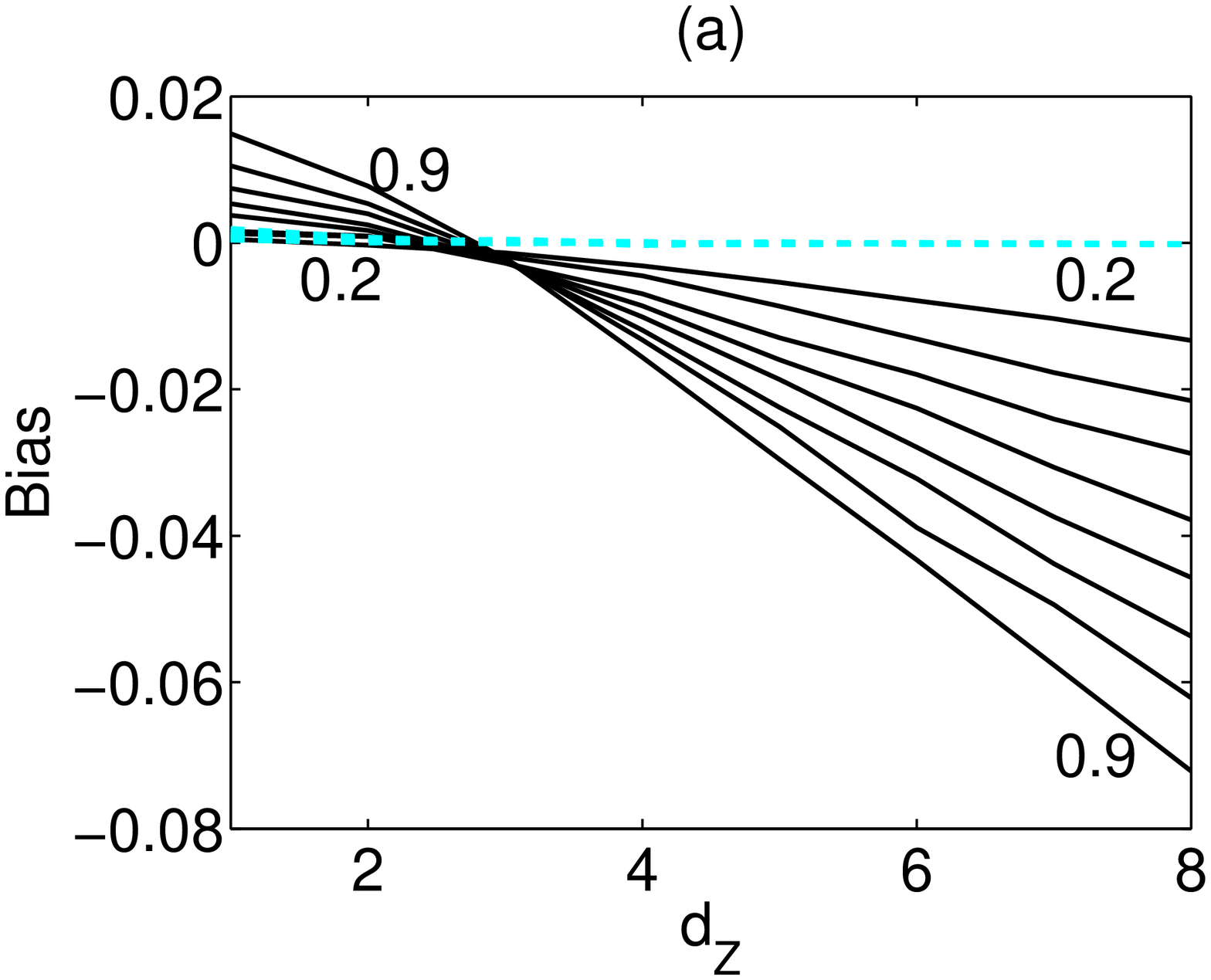} 
\includegraphics[height=3.3cm]{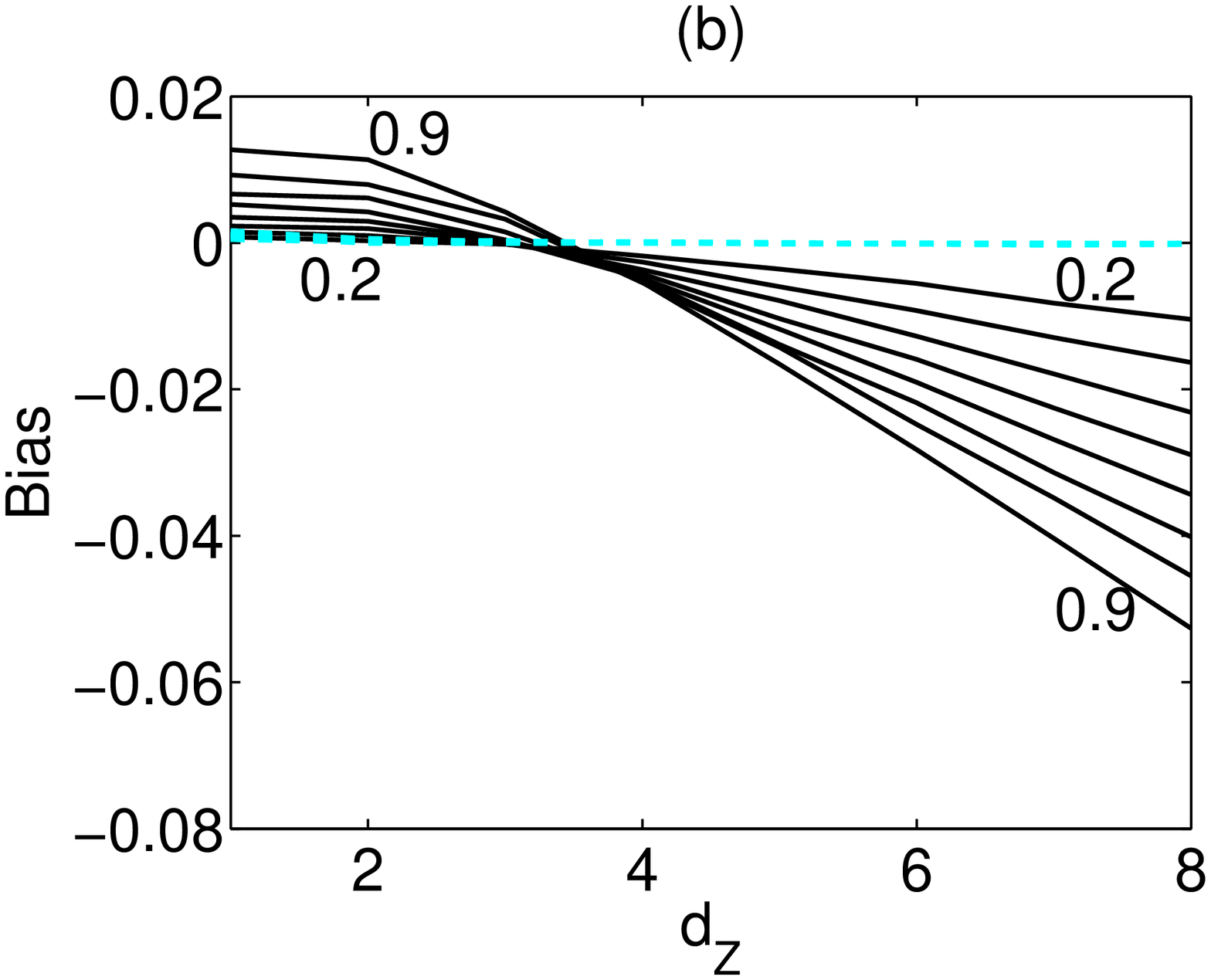}}}
\centerline{\hbox{\includegraphics[height=3.3cm]{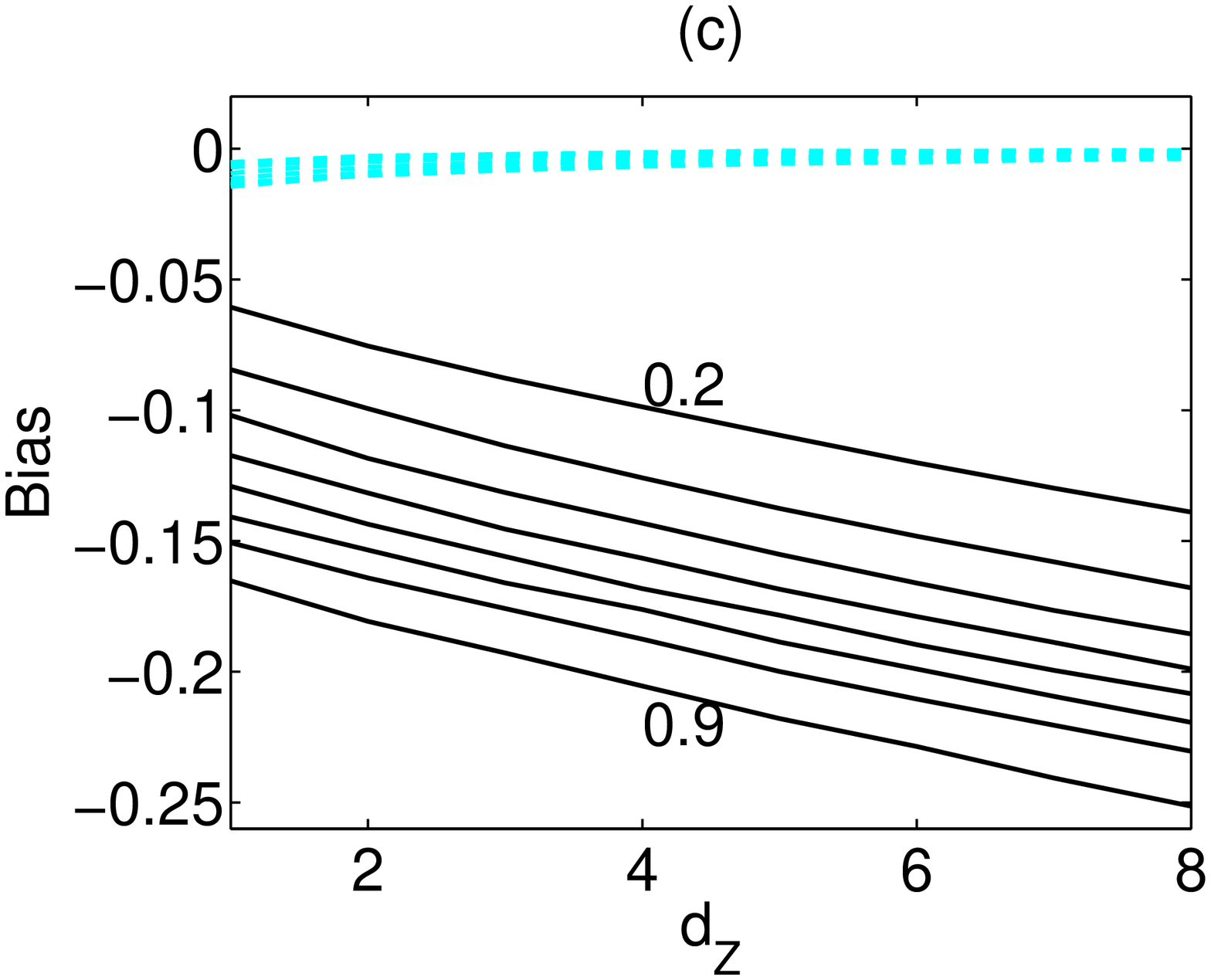} 
\includegraphics[height=3.3cm]{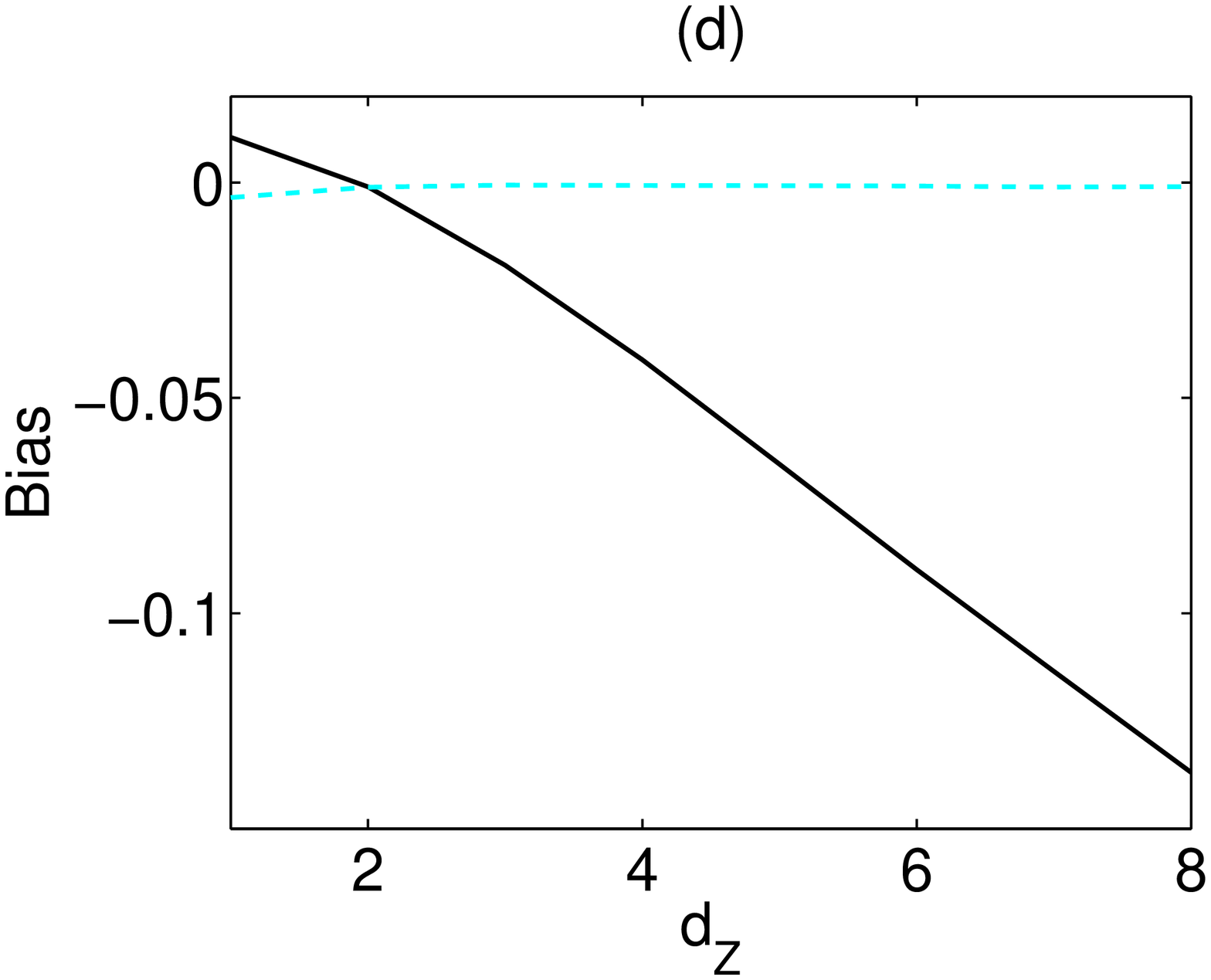}}}
\caption{(Color online) Bias of mutual information with black solid line and conditional mutual information with gray dashed line (cyan on line) for
Gaussian time series with (a) $d_X=d_Y=1$ and $d_Z=1,\ldots,8$, $N=4096$, (b) $d_X=d_Y=1$ and $d_Z=1,\ldots,8$, $N=8192$, (c) $d_X=8,d_Y=1$ and
$d_Z=1,\ldots,8$, $N=4096$. Lines correspond to different correlation coefficient $r=0.2,\ldots,0.9$ with step 0.1. In (d) we have the case of
varying correlation coefficients with $d_X=d_Y=1$ and $d_Z=1,\ldots,8$, $N=4096$.} \label{fig:MCMbias}
\end{figure}

The theoretical values for $I(X;(Y,Z))$ and  $I(X;Y \mid Z)$ can be computed with the
help of the expression of entropy for a multivariate Gaussian process $X$ with unit variance
\cite{Cover06}
\begin{equation*}
H(X) = {1 \over 2}\log\left((2\pi e)^d |R|\right),\label{TheorMIG}
\end{equation*}
with $d$ the dimension of the variable $X$ and $|R|$ the determinant of the correlation
matrix $R$ of $X$. We estimate the average MI and CMI of 1000 simulations for each case
(and different $d_Z$ and $r$) and compute the mean of the bias terms for $I(X;(Y,Z))$ and
$I(X;Y \mid Z)$, given as $<\hat I>-I_{\mathrm{theoretical}}$. In Fig.~\ref{fig:MCMbias}
we give the bias for all cases.
We see that the bias of CMI is constantly close to zero,
while for MI it is large and can be both negative and positive depending on the dimension
and correlation of the variables.
Figures~\ref{fig:MCMbias}(a and b) show that as sample size increases MI bias barely
reduces. In Fig.~\ref{fig:MCMbias}(c) where $d_X=8$ the bias is much larger for MI, and
in Fig.~\ref{fig:MCMbias}(d) the case of varying correlation coefficients shows the same
results. The number of neighbors used for the estimation of $I(X;(Y,Z))$ and $I(X;Y \mid
Z)$ in this example is set to 10 as in all the Monte Carlo simulations in our study.
This choice was decided after a pilot study on a wide range of $K$ values (from 5 to 30)
showed that though the mutual information values changed with $K$ the embedding
vector forms were stable. We note that for a small time series length, e.g. $N=512$, a
smaller value of $K$ may be more appropriate but for consistency purposes we use $K=10$
for all $N$.

The analysis above shows that CMI is a better information measure to use in our embedding
scheme than MI. Denoting $X=\mathbf{x}_{F}$, $Y=x_j$ and
$Z=\mathbf{b}_{j-1}$, from Eq.~\eqref{ivlach:e5} CMI is related to MI as
\begin{eqnarray}\label{ivlach:e6}
I(\mathbf{x}_{F};x_j \mid \mathbf{b}_{j-1})=I(\mathbf{x}_{F};(x_j,\mathbf{b}_{j-1}))-
I(\mathbf{x}_{F};\mathbf{b}_{j-1}).
\end{eqnarray}
Thus instead of using $I(\mathbf{x}_{F};(x_j,\mathbf{b}_{j-1}))$ in the information
criterion for embedding of Eq.~\eqref{ch6:simplecriterion}, we can use
$I(\mathbf{x}_{F};x_j \mid \mathbf{b}_{j-1})$. We refer to the two criteria as
\begin{eqnarray*}
\mathrm{I}_0&:&\max\limits_{x_j}\left\{I\big(\mathbf{x}_{F};\big(x_j,\mathbf{b}_{j-1}\big)\big)\right\},\\
\mathrm{I}_1&:&\max\limits_{x_j}\left\{I\big(\mathbf{x}_{F};x_j \mid \mathbf{b}_{j-1}\big)\right\}.
\end{eqnarray*}

Theoretically, both criteria are equivalent as the term $I(\mathbf{x}_{F};\mathbf{b}_{j-1})$
in Eq.~\eqref{ivlach:e6} is constant at step $j$ with respect to $x_j$. However,
in practice this is not true, since the estimation of $I(\mathbf{x}_{F};\mathbf{b}_{j-1})$
is done by the formula in Eq.\eqref{ivlach:e8} that depends on $x_j$. Due to
the lower bias in the estimation of CMI, as shown above, we expect $\mathrm{I}_1$ criterion
to perform better.

Another serious effect of the bias in the estimation of MI is on the stopping criterion
of the proposed state space reconstruction in Eq.~\eqref{ch6:threshold}
(common for both $\mathrm{I}_0$ and $\mathrm{I}_1$ criteria).
Let us suppose that the two terms $I\big(\mathbf{x}_{F};\mathbf{b}_{j-1}\big)$ and $I\big(\mathbf{x}_{F};\mathbf{b}_{j}\big)$ in the stopping criterion are
estimated independently. The difference in dimension by one of $\mathbf{b}_{j-1}$
and $\mathbf{b}_{j}$ regards a different bias in the two MI terms.
The correlation structure may also contribute to the bias
but at a lesser extent as it does not change a lot by the inclusion of a single component.
The bias can even be negative, and moreover,
the bias for $I\big(\mathbf{x}_{F};\mathbf{b}_{j}\big)$ may be larger in
magnitude than for $I\big(\mathbf{x}_{F};\mathbf{b}_{j-1}\big)$,
so that $\hat I\big(\mathbf{x}_{F};\mathbf{b}_{j-1}\big)> \hat
I\big(\mathbf{x}_{F};\mathbf{b}_{j}\big)$ may occur, which cannot happen in theory.
This theoretically impossible result was observed in simulations.

In order to balance the bias due to dimension change and correlation structure
we estimate $I\big(\mathbf{x}_{F};\mathbf{b}_{j-1}\big)$ using a
projection of the space regarding $I\big(\mathbf{x}_{F};(x_j,\mathbf{b}_{j-1})\big)$,
as given in Eq.\eqref{ivlach:e8}. As we have seen, CMI has near zero
bias, and thus the bias of $\hat I_{p}\big(\mathbf{x}_{F};\mathbf{b}_{j-1}\big)$
should be approximately equal to the bias of $\hat
I\big(\mathbf{x}_{F};(x_j,\mathbf{b}_{j-1})\big)$ (estimated by Eq.~\eqref{ivlach:e7}), making the two terms better comparable in the stopping criterion.

To demonstrate the two non-uniform multivariate embedding schemes we give an example for
the $x$ and $z$ variables of the Lorenz system \cite{Lorenz63}
\begin{eqnarray*}
\dot x(t)&=&10(y(t)-x(t))\\
\dot y(t)&=&x(t)(28-z(t))-y(t)\\
\nonumber \dot z(t)&=&x(t)y(t)-8/3 z(t),
\end{eqnarray*}
contaminated with observational Gausian white noise added to each variable with a
standard deviation 10\% of the standard deviation of the variable. The time series length
is 10000 and the sampling time is $\tau_s=0.05$ sec. We are interested in predicting the $x$
variable and set $\mathbf{x}_{F}=(x_{n+1},\ldots,x_{n+5})$. We set $L_{x} = L_{z}=25$ and
use the value $A=0.95$ for the stopping threshold value.

\begin{figure}[htb]
\centerline{\hbox{\includegraphics[height=3.3cm]{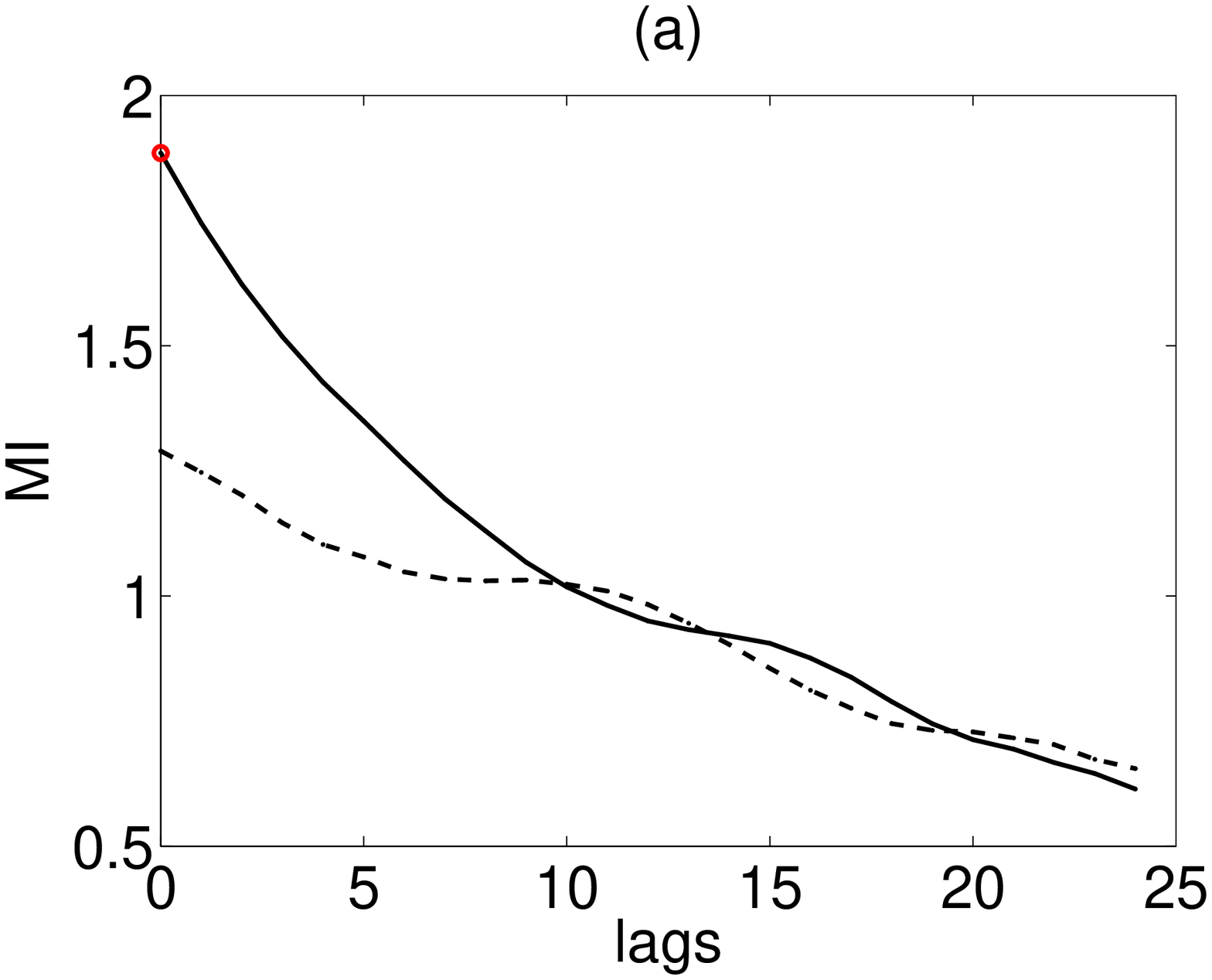}
\includegraphics[height=3.3cm]{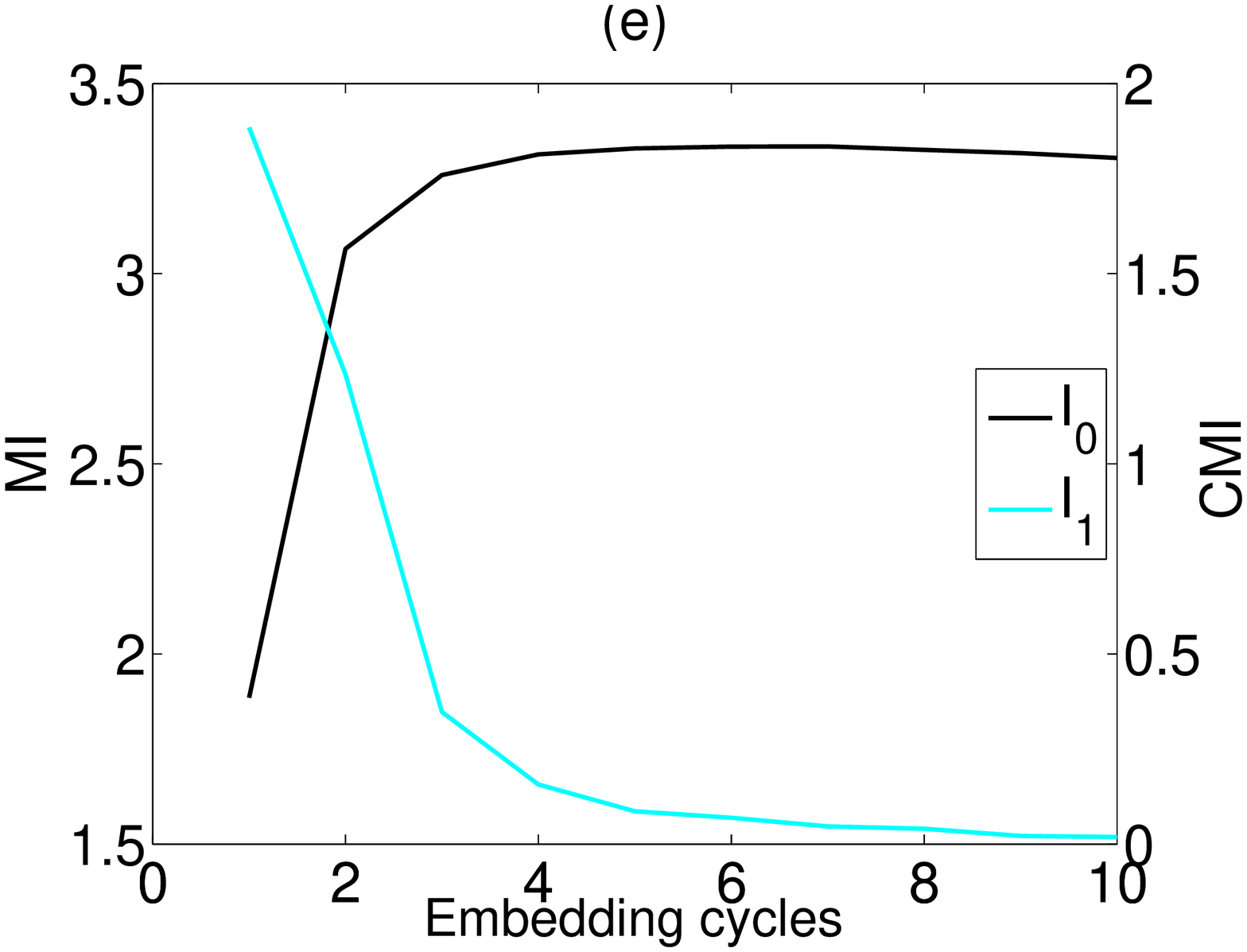}}}
\centerline{\hbox{\includegraphics[height=3.3cm]{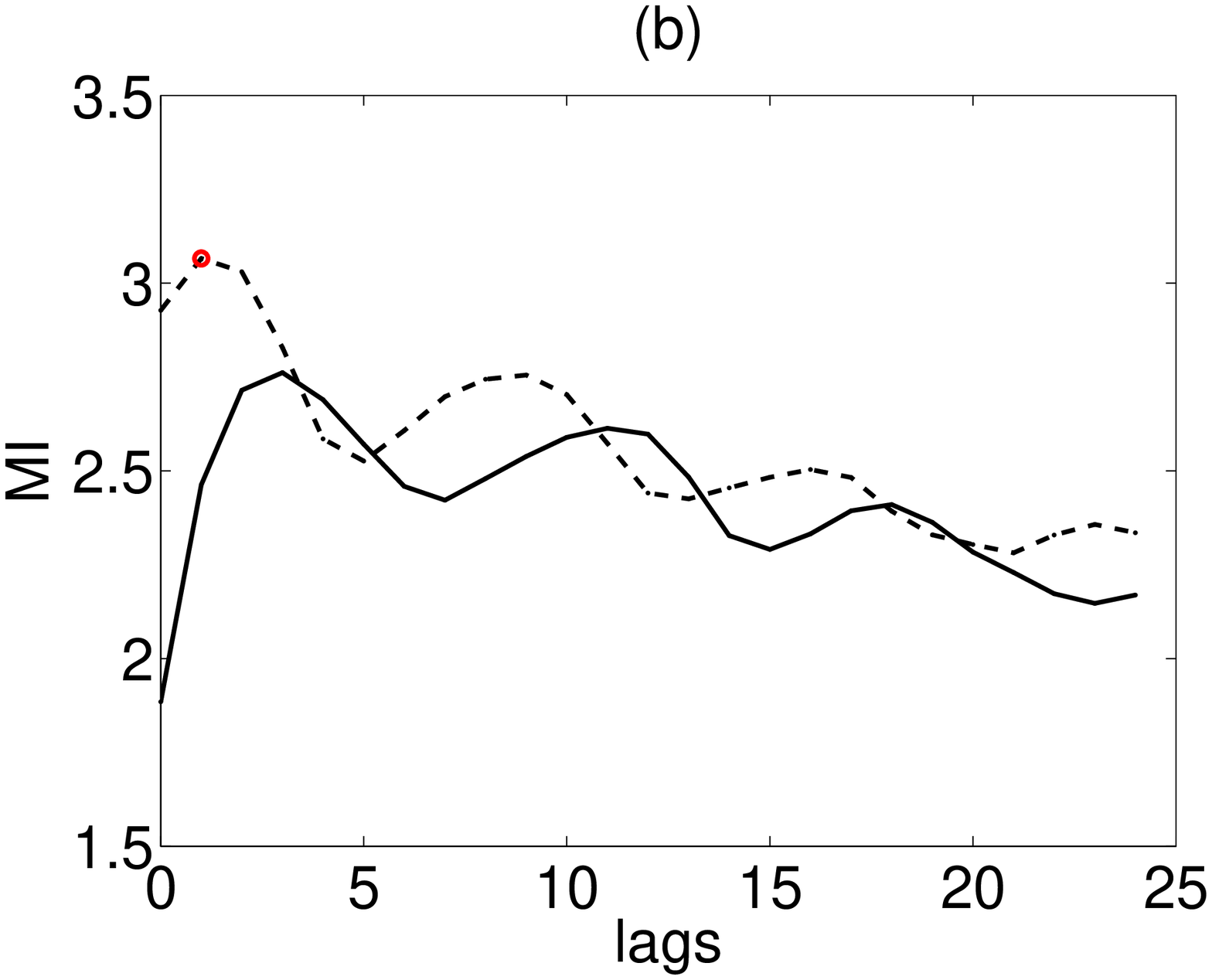}
\includegraphics[height=3.3cm]{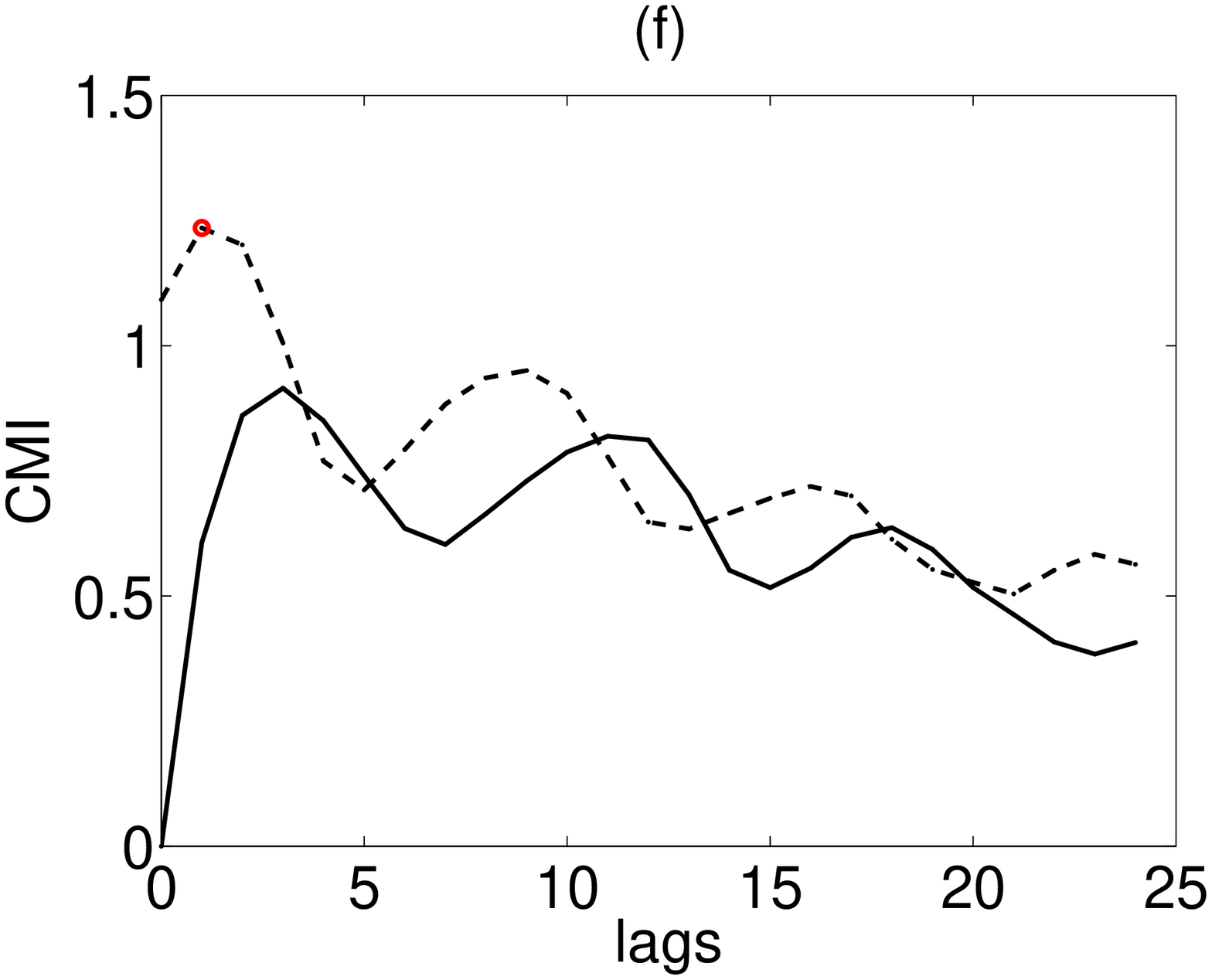}}}
\centerline{\hbox{\includegraphics[height=3.3cm]{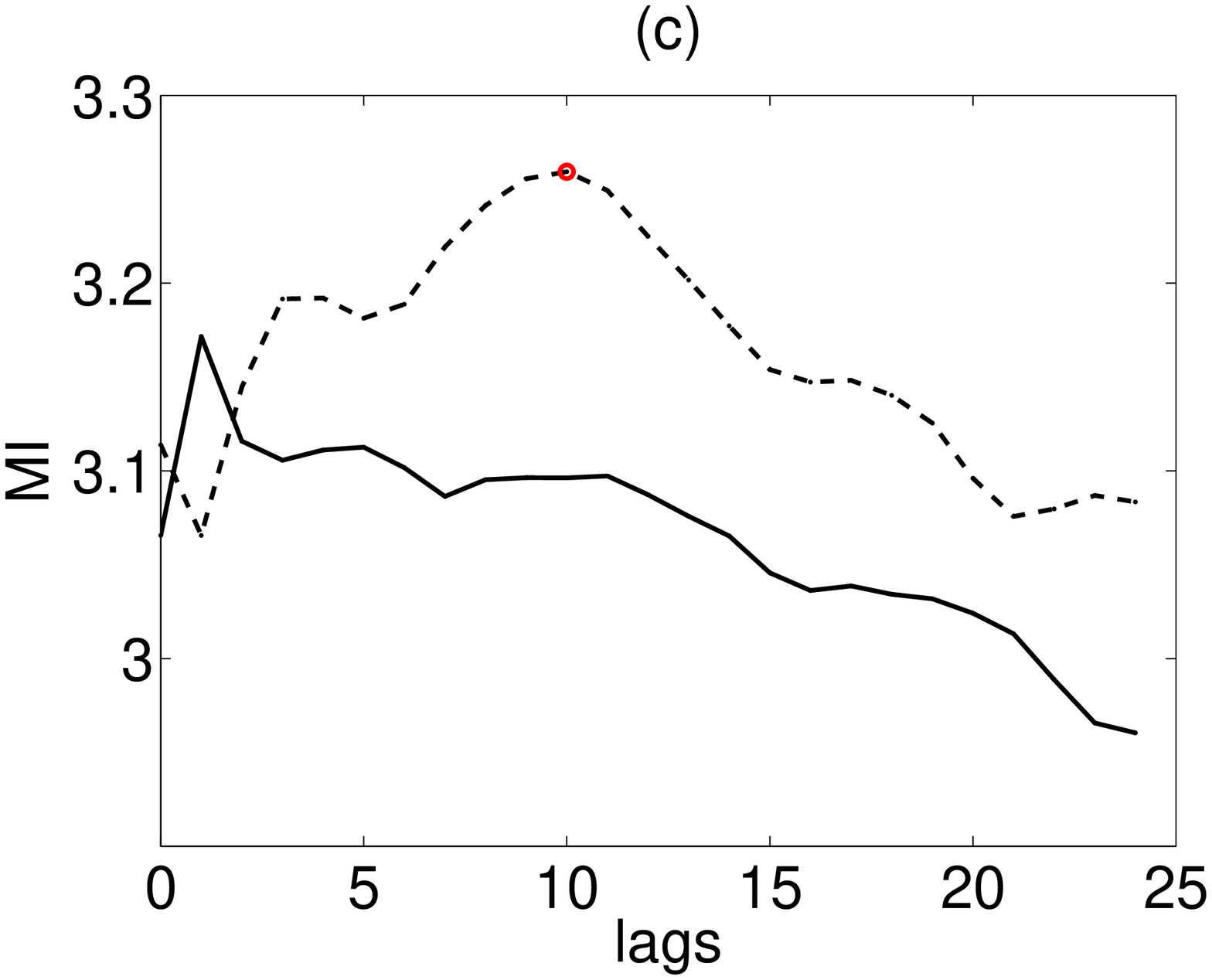}
\includegraphics[height=3.3cm]{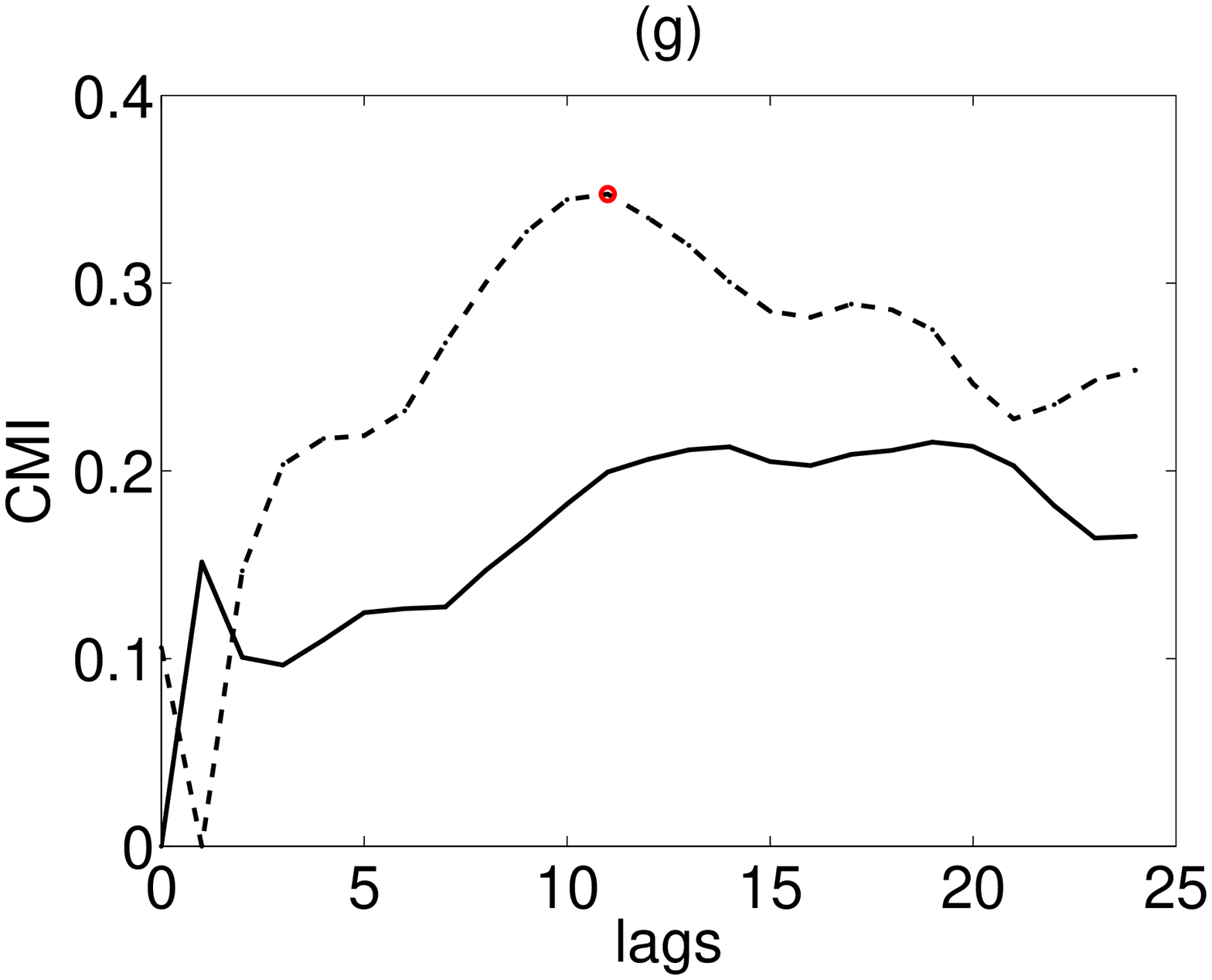}}}
\centerline{\hbox{\includegraphics[height=3.3cm]{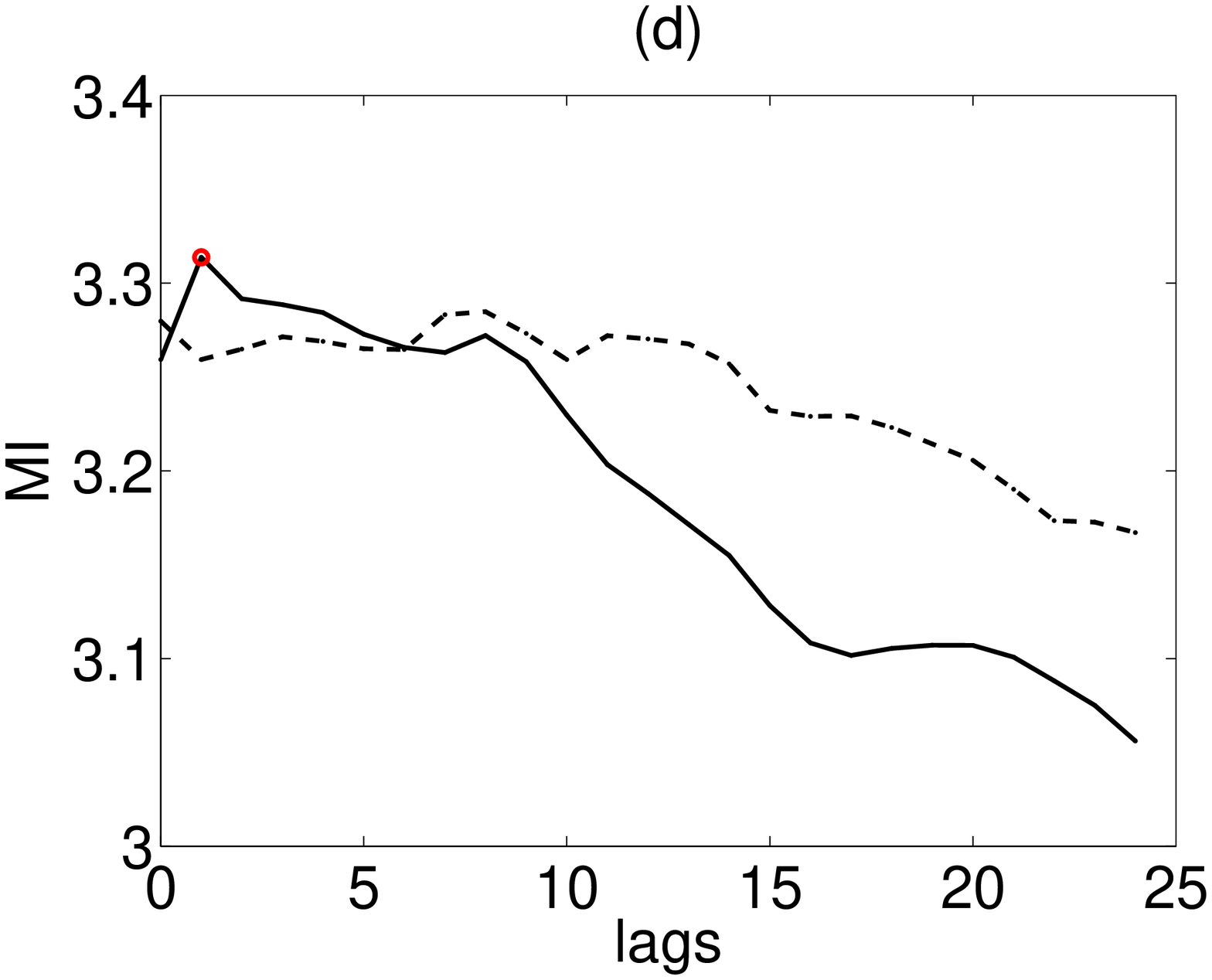}
\includegraphics[height=3.3cm]{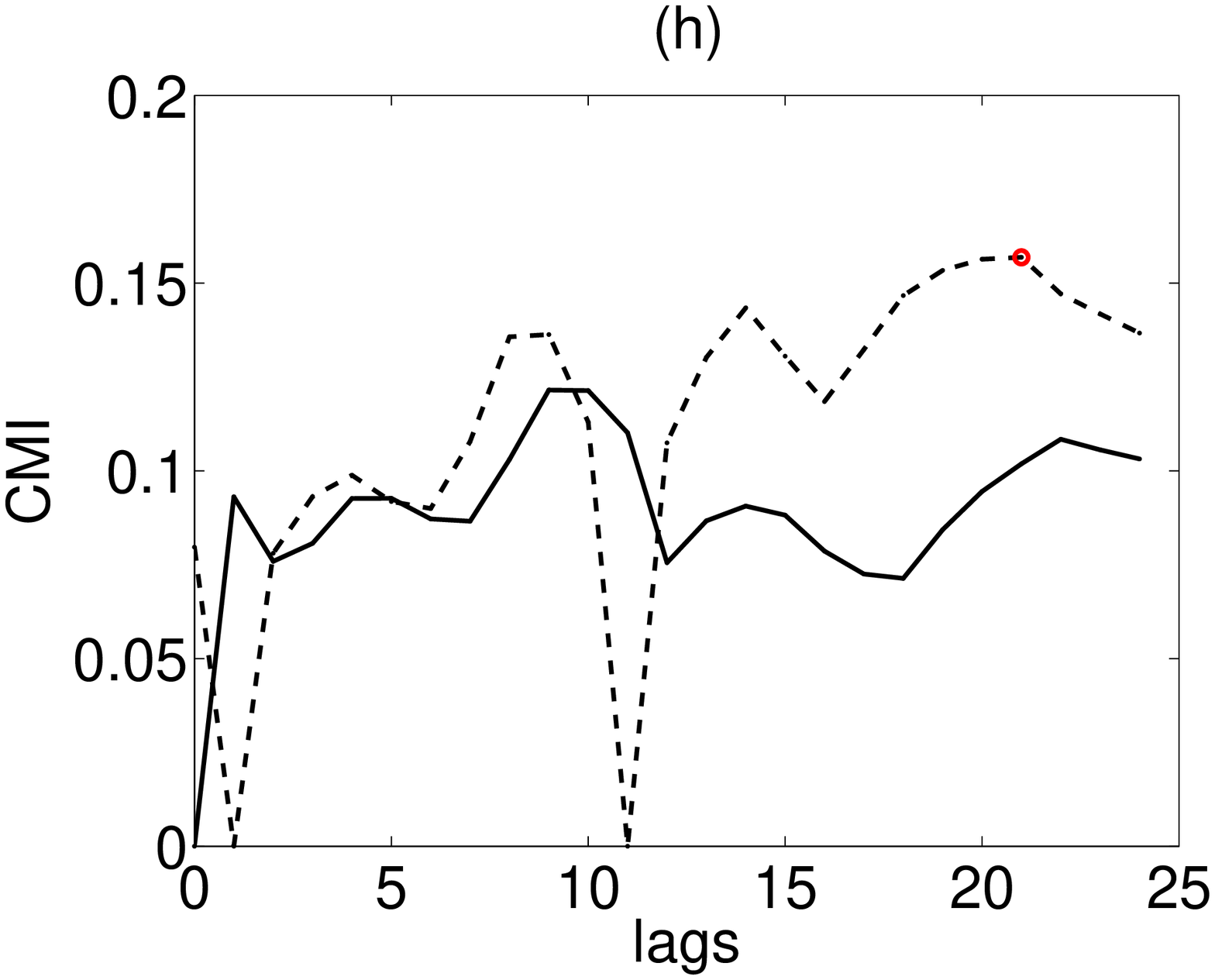}}}
\caption{(Color online)(a)-(d) The first four embedding cycles of $\mathrm{I_0}$ for the reconstruction aiming to explain Lorenz's $x$ variable from
$x$ and $z$ variables. Each panel shows the mutual information for all candidate components at each iteration, where the solid line represents lags
of $x$, the dotted line represents lags of $z$ and the selected lag is highlighted. The selected lag in panel (d) is not included in the embedding
vector because the ratio exceeds the threshold value. Panels (f)-(h) show the corresponding to (b)-(d) embedding cycles using $\mathrm{I_1}$ and
showing the conditional mutual information. The first embedding cycle is omitted because it is identical to (a). Panel (e) has the values of mutual
information (criterion $\mathrm{I_0}$ - black line) and conditional mutual information (criterion $\mathrm{I_1}$ - grey line, cyan online) for 10
embedding cycles, with the left and right vertical axes corresponding to the values of MI and CMI, respectively.} \label{fig:Lxz}
\end{figure}

For the given threshold, the obtained embedding vectors for the $\mathrm{I_0}$ criterion
have the form $(x_{n},z_{n-1},z_{n-10})$, while $\mathrm{I_1}$ gives
$(x_{n},z_{n-1},z_{n-11})$. In Fig.~\ref{fig:Lxz}, the values of the respective MI and
CMI are given for the first 4 embedding cycles (an iteration that results in the addition
of a component in the embedding vector), along with the selected component at each cycle.

The first embedding cycle is identical for both $\mathrm{I}_0$ and $\mathrm{I}_1$ as the
conditioning vector is empty, and the forms of MI and CMI are very similar at the second
embedding cycle. They differ a bit at the third embedding cycle, which leads to the
selection of different components ($z_{n-10}$ for $\mathrm{I_0}$ and $z_{n-11}$ for
$\mathrm{I_1}$), and consequently the fourth embedding cycle gives completely different
components. In Fig.~\ref{fig:Lxz}(e) we give the MI and CMI for 10 iterations of the
embedding procedure.

MI increases to a maximum value with the embedding cycles and
then decreases, only slightly, due to the bias. CMI decreases gradually and monotonically
to zero, which means that the new components contribute gradually less.

\subsection{Estimation of information transfer using non-uniform multivariate embedding}

Non-uniform multivariate embedding can be used for measuring information flow between coupled systems. The main approach for estimating information
transfer is based on the idea of Granger causality \citep{Granger80}. Simply put,
having measurements from $x$ and $y$, a causal relationship from a time series
$x$ to another $y$, can be identified if the prediction of $y$ gets better
by including also values of $x$.
One way to implement this idea is to find the best model for predicting $y$ from
its previous values and the best model for predicting $y$ from previous values of
$y$ and $x$, and compare the two predictions. If the mixed prediction that uses
both $y$ and $x$ is better, then there is causal relationship from $x$ to $y$.

An interesting fact is that the idea of Granger causality contradicts Taken's embedding
theorem, according to which for explaining the dynamics of $y$ (and consequently
predicting its values) a time delay reconstruction from components of $y$ is sufficient,
and therefore $x$ values are not needed. As a matter of fact, in \cite{Faes08} a
measure related to Granger causality was used in a case of strongly unidirectionally
coupled Henon maps and the causality relation could not be detected. This was because
both the best univariate model and the best mixed model gave approximately equally good
predictions. Our approach of detecting causality aims at rendering this problem.
Simply, if the embedding procedure for prediction of $y$ results
in embedding vectors that contain components of $x$, this means that information from $x$
is transferred to $y$.

Depending on the coupling strength and the nature of the
variables, some adjustment of the parameters may be needed. If the coupling strength is
quite small the stopping criterion threshold may need to be increased to values over
0.95, and in the case of flows or time series with different complexity, different time horizons
spanned by $\mathbf{x}_{F}$ should be tested.

Further, we introduce two ways to quantify the coupling strength if the embedding procedure results in a state space of mixed components, i.e. from
$x$ and $y$. Suppose that the embedding vector aiming to explain $\mathbf{x}_{F}$ that contains only future values of $y$ (e.g.
$\mathbf{x}_{F}=(y_{n+1})$) has the form
$$\mathbf{x}_n=(x_{n-l_{11}},x_{n-l_{12}},\ldots,x_{n-l_{1m_1}},y_{n-{l_{21}}},\ldots,y_{n-l_{2m_2}}).$$
We estimate one-step ahead prediction with local model on these embedding vectors. For each target point in the set of all reconstructed points, the
prediction is simply the mapping of its nearest neighbor found from all other points in the set, and we compute the normalized root mean square error
(NRMSE) $E$.
\begin{equation*}
E_{\mathbf{x}}=\sqrt{{\sum\limits_{n}( y_{n+1}-\hat{y}_{n+1})^{2} \over \sum\limits_{n}( y_{n+1}-\bar{y})^{2}}}, \label{eqch6:NRMSE}
\end{equation*}
where $\hat{y}_{n+1}$ is the in-sample one step ahead prediction at time $n$ and
$\bar{y}$ is the sample mean. Then we consider the reduced embedding vectors using
components only from $x$ or $y$
$$\mathbf{x}_n^1=(x_{n-l_{11}},\ldots,x_{n-l_{1m_1}}),\quad \mathbf{x}_n^2=(y_{n-l_{21}},\ldots,y_{n-l_{2m_2}}).$$
We compute the $E$ of local model fit of $y_{n+1}$ using the projected embedding vectors $\mathbf{x}_n^2$, $E_{\mathbf{x}^2}$, and obtain the measure
of coupling from $x$ to $y$ as
\begin{equation}
S_{X \rightarrow Y}=1-{E_{\mathbf{x}} \over E_{\mathbf{x}^2}}.
\end{equation}

If $x$ drives $y$ the omittance of components of $x$ (the use of $\mathbf{x}^2_n$ instead of $\mathbf{x}_n$), would increase $E$, so that
$E_{\mathbf{x}^2}>E_{\mathbf{x}}$, and thus $S$ would be positive. If there is no coupling, then components of $x$ are not useful and either they are
not included in $\mathbf{x}_n$ ($\mathbf{x}_n$=$\mathbf{x}^2_n$) or they do not contribute significantly in predicting $y$, so that
$E_{\mathbf{x}^2}\leq E_{\mathbf{x}}$. In any case, the result is that $S$ is not significant. In our study we varied the prediction to $T$-step
ahead depending on the system and form of $\mathbf{x}_{F}$.

We propose a similar measure to $S$ replacing $E$ with mutual information, giving
\begin{equation}
R_{X \rightarrow Y}=1-{I(\mathbf{x}_{F};\mathbf{x}^2) \over I(\mathbf{x}_{F};\mathbf{x})} ={I( \mathbf{x}_{F};\mathbf{x}^1 \mid \mathbf{x}^2) \over
I(\mathbf{x}_{F};\mathbf{x})}.
\end{equation}
$R_{X \rightarrow Y}$ measures the information of $y$ that is explained by components of
the embedding vectors that emanate from $x$, normalized against the total mutual
information in order to give a value between 0 and 1. The measures for the opposite
direction are defined accordingly.

\section{Simulations and results}\label{ivlach:sec4}

\subsection{Driven Henon}

We evaluate the embedding vectors derived by criteria $\mathrm{I}_0$ and $\mathrm{I}_1$ and
test our embedding scheme for coupling strength estimation on some coupled systems.
First is the driver-response Henon system \cite{Schiff96} given by
\begin{equation*}
\begin{array}{lll}
x_{n+1}&=&1.4-x_{n}^2+0.3x_{n-1}\\
y_{n+1}&=&1.4-\left(C y_n x_{n}+(1-C)y_n^2\right)+0.3y_{n-1}
\end{array}
\end{equation*}
The time series length is 4096 and the coupling strength takes values $C$=[0, 0.05, 0.1,
0.2, 0.3, 0.4, 0.5, 0.6]. The non-uniform multivariate embedding procedure is applied on
100 realizations for each $C$ and the embedding vector that has the largest frequency of
occurrence is selected. We set $L_{x} = L_{y}=5$, to account for all significant delays,
and $\mathbf{x}_{F}=(x_{n+1})$ for predicting $x$ and $\mathbf{x}_{F}=(y_{n+1})$ for $y$,
as one step ahead is sufficient to detect causal effects (e.g. see \cite{Faes08}). We let
the stopping criterion threshold $A=0.95$ as for maps small variations of $A$ do not
alter the form of the embedding vectors.

Across the 100 realizations for each $C$, a single embedding vector form was
found in both methods (but not necessarily the same for the two methods). For the prediction of $x$ both
$\mathrm{I}_0$ and $\mathrm{I}_1$ produced the vector $(x_n,x_{n-1})$ for all $C$, while
for the prediction of $y$ the two methods differed slightly: method $\mathrm{I}_0$ produced
$(y_n,y_{n-1})$ for $C=0, 0.05, 0.1$, $(x_{n-1},y_n,y_{n-1})$ for $C=0.2,0.3,0.4$ and
$(x_n,x_{n-1},y_n,y_{n-1})$ for $C=0.5, 0.6$, while  method $\mathrm{I}_1$ differed only for
$C=0.1$ selecting the form $(x_{n-1},y_n,y_{n-1})$, i.e., detecting the correct driving also
for weaker coupling strength.

\begin{figure*}[htb]
 \centerline{\includegraphics[height=4.1cm]{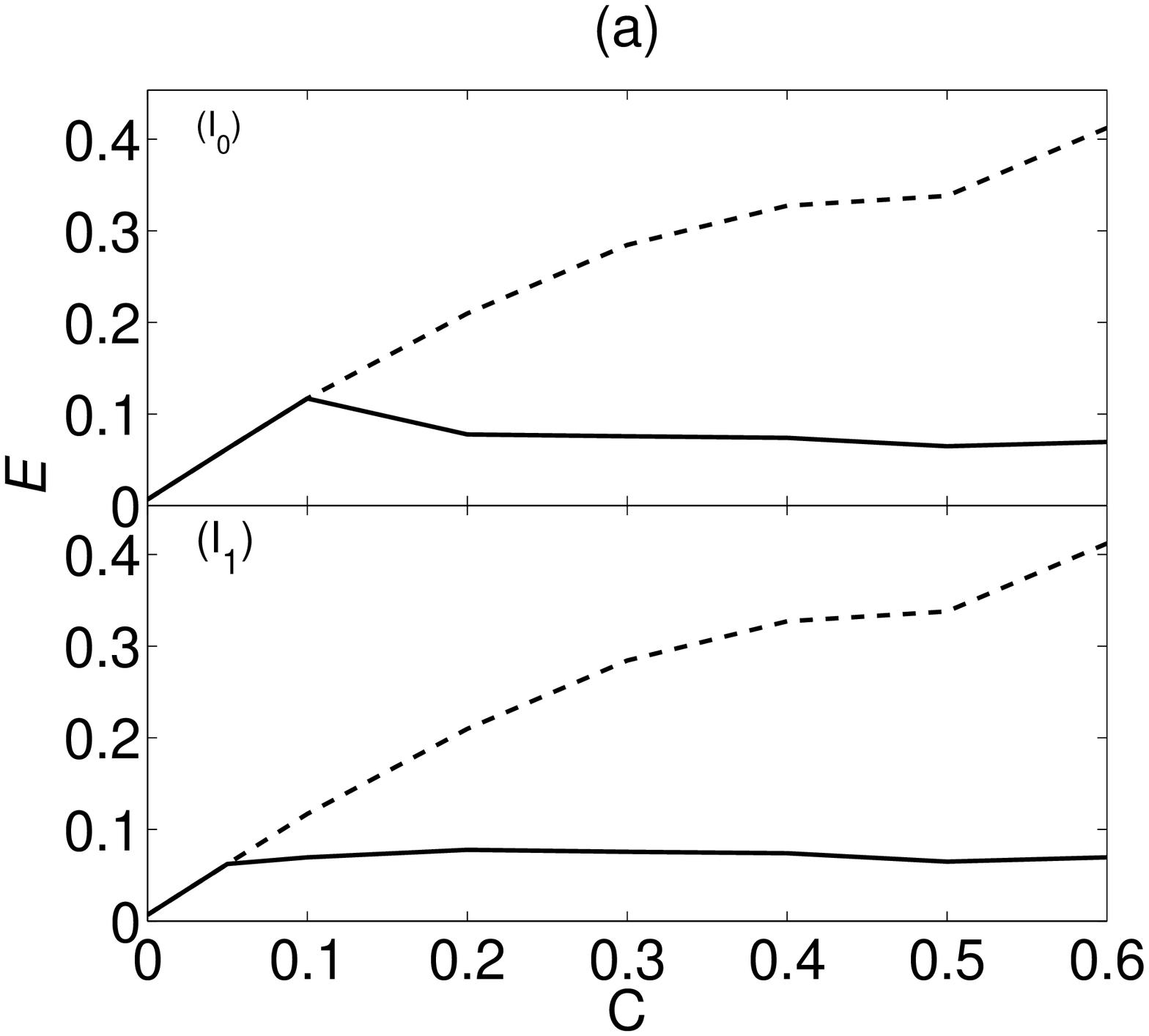}
\hspace{-0.3cm}
\includegraphics[height=4.1cm]{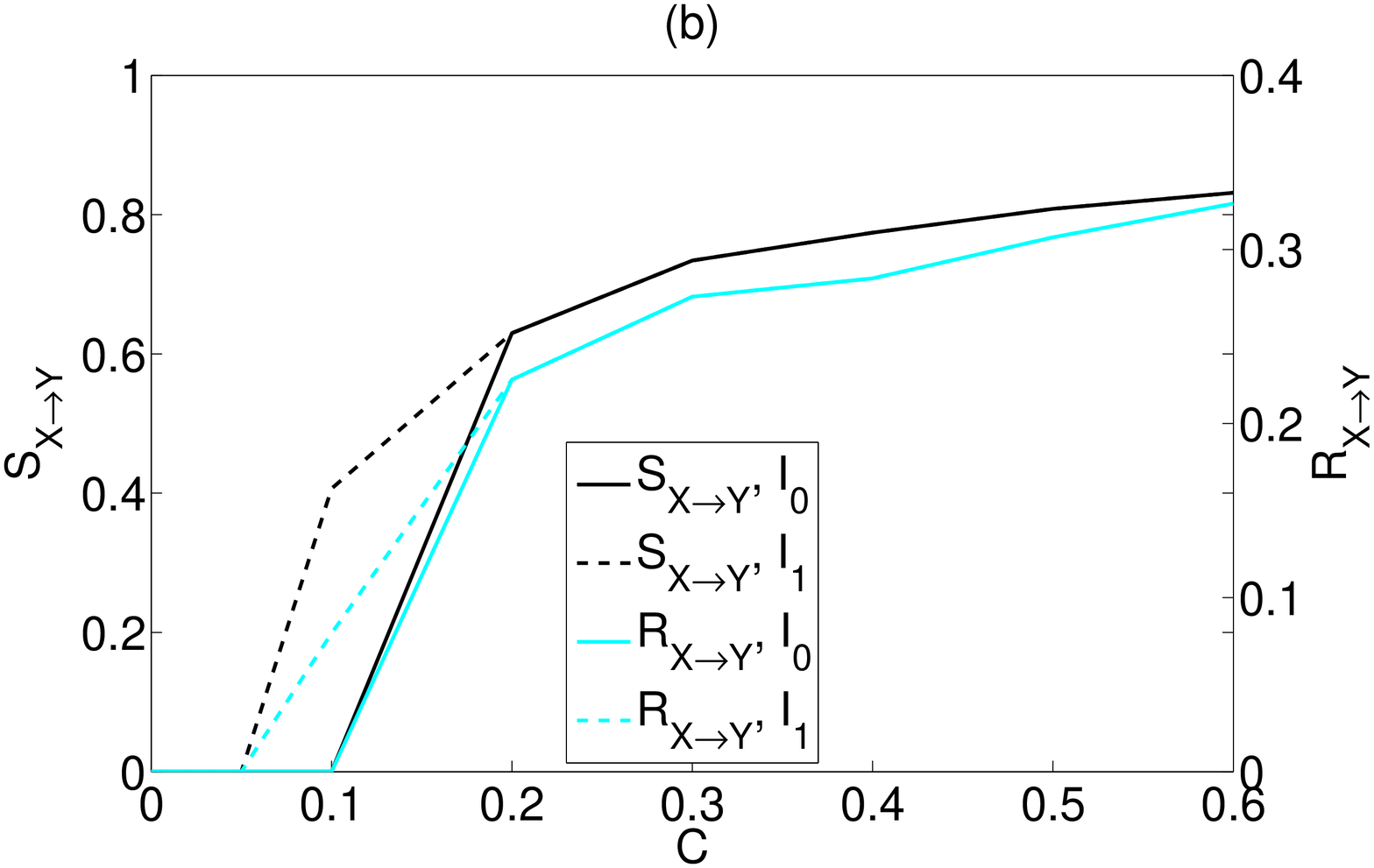}\hspace{0cm}\includegraphics[height=4.1cm]{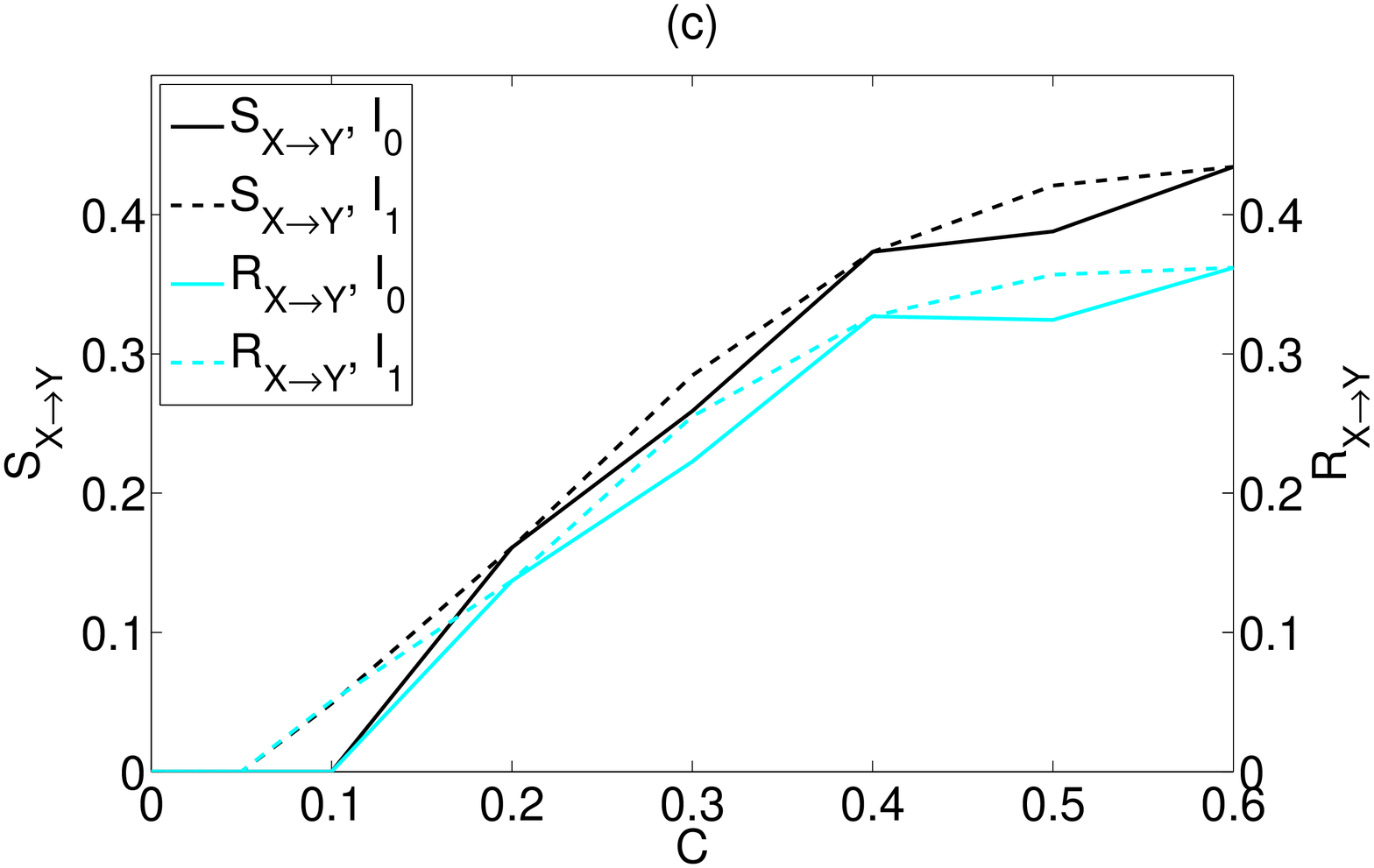}}
\caption{(Color online) (a) $E_{\mathbf{x}}$ (solid line) and $E_{\mathbf{x}^2}$ (dashed line) for the driven Henon $y$ across all $C$, where
embedding vectors are suggested by $\mathrm{I_0}$ in the upper panel and $\mathrm{I_1}$ in the lower. (b) $S_{X \rightarrow Y}$ and $R_{X \rightarrow
Y}$ for the same system across all $C$ and for both criteria (solid line for $\mathrm{I_0}$ and dashed for $\mathrm{I_1}$, with the left vertical
axis corresponding to values of $S$ and the right of $R$). (c) The same as (b) for unidirectionally coupled Henon maps with 20\% noise.}
\label{fig:CHN}
\end{figure*}

On the same 100 realizations the selected embedding vectors are used for the estimation of average values for $E_{\mathbf{x}}$, $E_{\mathbf{x}^2}$,
as well as the $S_{X \rightarrow Y}$ and $R_{X \rightarrow Y}$ coupling measures and are shown in Fig.~\ref{fig:CHN}.
The results on the measures are in agreement with the form of the embedding vectors.
The monotonic rise of
the $S_{X \rightarrow Y}$ curve shows that as $C$ increases the contribution of $x$ in
predicting $y$ also increases. The use of criterion $\mathrm{I_1}$ shifts the detection
of coupling to smaller $C$. The results for the $R_{X \rightarrow Y}$ measure for both
criteria are similar to those for $S_{X \rightarrow Y}$.

We add 20\% observational noise to each variable ($x$ and $y$) and repeat the analysis.
Table \ref{tab:DRHenonNoise} shows the selected embedding vectors and their frequency.
\begin{table*}[htb]
\caption{\label{tab:DRHenonNoise}Embedding vectors and their frequency of occurrence for the system of unidirectionally coupled Henon maps with 20\%
noise.}
\begin{center}
\begin{tabular*}{1\textwidth}{@{\extracolsep{\fill}}lrrrr}
\hline\hline $C$&$\mathrm{I_0}$ $(x,y) \to x$&$\mathrm{I_0}$ $(x,y) \to y$&$\mathrm{I_1}$ $(x,y) \to x$&$\mathrm{I_1}$ $(x,y) \to y$\\\hline

0& $(x_n,x_{n-1},x_{n-2})$    100& $(y_n,y_{n-1},y_{n-2})$ 97 & $(x_n,x_{n-1},x_{n-2})$ 92& $(y_n,y_{n-1},y_{n-2})$ 91\\

0.05& $(x_n,x_{n-1},x_{n-2})$ 100& $(y_n,y_{n-1},y_{n-2})$ 97& $(x_n,x_{n-1},x_{n-2})$ 92& $(y_n,y_{n-1},y_{n-2})$ 80\\

0.1& $(x_n,x_{n-1},x_{n-2})$  100& $(y_n,y_{n-1},y_{n-2})$ 53& $(x_n,x_{n-1},x_{n-2})$ 92& $(x_{n-1},y_n,y_{n-1},y_{n-2})$ 42\\

&&                                 $(y_n,y_{n-1})$ 22        &                           &  $(y_n,y_{n-1},y_{n-2})$ 31\\

0.2& $(x_n,x_{n-1},x_{n-2})$  100& $(x_{n-1},y_n,y_{n-1},y_{n-2})$ 52 & $(x_n,x_{n-1},x_{n-2})$ 92& $(x_{n-1},y_n,y_{n-1},y_{n-2})$ 45\\

 &&                                $(x_{n-1},y_n,y_{n-1})$ 33          &                          & $(x_{n-1},y_n,y_{n-1})$ 28\\

0.3& $(x_n,x_{n-1},x_{n-2})$  100& $(x_n,y_n,y_{n-1})$ 55 & $(x_n,x_{n-1},x_{n-2})$ 92& $(x_n,x_{n-2},y_n,y_{n-1})$ 47\\

                                 &&$(x_{n-1},y_n,y_{n-1})$ 19 &                       &$(x_n,y_n,y_{n-1})$ 27\\

0.4& $(x_n,x_{n-1},x_{n-2})$  100& $(x_n,x_{n-2},y_n,y_{n-1})$ 34 & $(x_n,x_{n-1},x_{n-2})$ 92& $(x_n,x_{n-2},y_n,y_{n-1})$ 79\\

&&                                 $(x_n,y_n,y_{n-1},y_{n-2})$ 27&                            & $(x_n,y_n,y_{n-1},y_{n-2})$ 10\\

0.5& $(x_n,x_{n-1},x_{n-2})$  100& $(x_n,y_n,y_{n-1})$ 52 & $(x_n,x_{n-1},x_{n-2})$ 92& $(x_n,x_{n-2},y_n,y_{n-1})$ 53\\

                                 && $(x_n,x_{n-2},y_n,y_{n-1})$ 19 &                  & $(x_n,y_n,y_{n-1},y_{n-2})$ 30\\

0.6& $(x_n,x_{n-1},x_{n-2})$  100& $(x_n,y_n,y_{n-1})$ 89& $(x_n,x_{n-1},x_{n-2})$ 92& $(x_n,y_n,y_{n-1})$ 85\\\hline\hline
\end{tabular*}
\end{center}
\end{table*}
When the frequency of the most frequently selected vector is small, also the second most frequently selected vector is shown. Careful examination of
the second most frequently selected vector form shows that these vectors are either the most selected vector forms with omittance of a component, or
with replacement of a component. The omittance is due to the threshold value that may be strict and in conjunction with the presence of noise at
cases does not allow for another component inclusion. The replacement is due to almost equivalent information in the regarded components, so that
due to noise perturbation either of them is selected ``randomly''. Figure~\ref{fig:CHN}(c) shows $S_{X \rightarrow Y}$ and $R_{X \rightarrow Y}$
calculated for the most frequently selected embedding vectors.

The presence of noise changes the embedding vectors and decreases $S_{X \rightarrow Y}$
values, but the information transfer results are similar to the noise-free case. As shown
in Fig.~\ref{fig:CHN}(c), $R_{X \rightarrow Y}$ works also better for criterion
$\mathrm{I_1}$ than for $\mathrm{I_0}$, where for the latter $R_{X \rightarrow Y}$ does
not increase monotonically with $C$.

\subsection{Bidirectionally coupled Henon maps}

The next system is for two bidirectionally coupled identical Henon maps
\cite{Wiesenfeldt01} given by the equations
\begin{equation*}\label{eqap:bidchenonmap}
\begin{array}{lll}
x_{n+1}&=&1.4-x_{n}^2+0.3x_{n-1}+C_2\left(x_{n}^2-y_n^2\right)\\
y_{n+1}&=&1.4-y_{n}^2+0.3y_{n-1}+C_1\left(y_{n}^2-x_n^2\right).
\end{array}
\end{equation*}
$C_1$ is the strength of $x$ driving $y$ and $C_2$ the opposite. We apply only
$\mathrm{I_1}$ criterion for the embedding vector selection and set $L_{x} = L_{y}=5$,
$N=4096$, $\mathbf{x}_{F}=(y_{n+1})$ for $y$ and $\mathbf{x}_{F}=(x_{n+1})$ for $x$. The
results for some specific coupling strength values are given in the second and third
column of Table~\ref{tab:BDHENON}.
\begin{table*}[htb]
\caption{\label{tab:BDHENON}Embedding vectors and their frequency of occurrence for
bidirectionally coupled Henon maps using criterion $\mathrm{I_1}$, along with $S$ and $R$
values for the most frequently selected vectors.}
\begin{center}
\begin{tabular*}{1\textwidth}{@{\extracolsep{\fill}}lrrrrrr}
\hline\hline [$C_1,C_2$]&$\mathrm{I_1}$ $(x,y) \to x$&$\mathrm{I_1}$ $(x,y) \to y$&$S_{Y
\rightarrow X}$&$S_{X \rightarrow Y}$&$R_{Y \rightarrow X}$&$R_{X \rightarrow Y}$\\\hline

[0.05, 0.05]     & $(x_n,x_{n-1},y_n)$ 97   & $(y_n,y_{n-1},x_n)$ 99 & -0.102 &  -0.104  &  0.055   & 0.056\\

[0.1, 0.1]       & $(x_n,x_{n-1},y_n)$ 52  &  $(y_n,y_{n-1},x_n)$ 49 & 0.322  &  0.328 &   0.148  & 0.148\\

                & $(x_n,x_{n-2},y_n)$ 37  &  $(y_n,y_{n-2},x_n)$ 38 &        &          &          &\\

[0.1, 0.05]    &$(x_n,x_{n-1},y_n)$ 97    &$(y_n,y_{n-1},x_n)$ 100&  -0.151 &   0.407&    0.055&    0.158\\

[0.15, 0.05]     & $(x_n,x_{n-1},y_n)$ 100  &$(y_n,y_{n-1},x_n)$ 100 &   -0.140 &   0.601&    0.061&    0.227\\

[0.2, 0.05]      &  $(x_n,x_{n-1},y_n)$ 100  &$(y_n,y_{n-1},x_n,x_{n-1})$ 54  &  -0.080  &  0.639  &  0.066  &  0.268\\

                &                           &$(y_n,y_{n-1},x_n,x_{n-2})$ 16 &        &          &          &\\\hline\hline

\end{tabular*}
\end{center}
\end{table*}
For equal coupling strengths, there is exact
interchange of $x$ and $y$ in the embedding vectors. For the last case with $C_1=0.2$ and
$C_2=0.05$ we observe that when predicting $x$ the embedding vectors are smaller than
when predicting $y$, as we would expect. The fact that for all cases, except the last one, the embedding
vectors have the same form, does not allow us to derive conclusions about the individual
coupling strengths and we turn to quantitative results of $S$ and $R$ measures.

The negative values of $S_{Y \rightarrow X}$ (see Table~\ref{tab:BDHENON}) show that the coupling cannot be detected by $S$ when it is very
weak ($C_2=0.05$) and the same holds for $S_{X \rightarrow Y}$ when $C_1=0.05$. When
there is substantial coupling of equal strength in both directions ($C_1=C_2=0.1$), $S_{X
\rightarrow Y}$ and $S_{Y \rightarrow X}$ take equally large values. When coupling
increases only in one direction, as for $C_2=0.05$ and $C_1=0.1,015,0.2$, the respective
$S_{X \rightarrow Y}$ measure increases as well. For weak coupling, $R$ performs better
than $S$ giving values slightly over zero and capturing the weak relationship between the
variables (see Table~\ref{tab:BDHENON}). As coupling increases so does the values of $R_{X \rightarrow Y}$ and the
difference between $R_{X \rightarrow Y}$ and $R_{Y \rightarrow X}$ shows the direction of
the strongest coupling.

\subsection{Unidirectionally coupled R\"{o}ssler--Lorenz}

The third system we study is the coupled R\"{o}ssler--Lorenz system \cite{Quyen99} given
by
\begin{eqnarray*}
\label{eqap:rosslerlorenzsystem}
\dot x_1(t)&=&6(-y_1(t)-z(t)_1)\\
\dot y_1(t)&=&6(x_1(t)+0.2 y_1(t)) \\
\dot z_1(t)&=&6(0.2+x_1(t)z_1(t)-5.7 z_1(t))\\
\dot x_2(t)&=&10(y_2(t)-x_2(t))\\
\dot y_2(t)&=&28x_2(t)-y_2(t)-x_2(t)z_2(t)+Cy_{1}(t)^2\\
\dot z_2(t)&=&-8/3 z_2(t)+x_2(t)y_2(t).
\end{eqnarray*}
We use the $y_1$ variable of the R\"{o}ssler system as the driving time series, which we
denote $x$, and $y_2$ of the Lorenz system for the response time series, which we denote
as $y$. The coupling strength takes values $C=[0, 0.5, 1, 1.5, 2, 3, 4]$.
We take $L_{x} = L_{y}=15$ to account for all significant delays.

First, we set $N=4096$ and apply the embedding procedure for $\mathbf{x}_{F}=(x_{n+1},x_{n+2},x_{n+3})$
for $x$ and $\mathbf{x}_{F}=(y_{n+1},y_{n+2},y_{n+3})$ for $y$, and let $A=0.95$ (discussion on
the choice of $\mathbf{x}_{F}$ and $A$ follows below). The most frequently selected
vectors from 100 realizations are given in Table~\ref{tab:DRroslor}.
\begin{table*}[htb]
\caption{\label{tab:DRroslor}Embedding vectors and their frequency of occurrence for the unidirectionally coupled R\"{o}ssler--Lorenz systems.}
\begin{center}
\begin{tabular*}{1\textwidth}{@{\extracolsep{\fill}}lrrrr}
\hline\hline $C$&$\mathrm{I_0}$ $(x,y) \to x$&$\mathrm{I_0}$ $(x,y) \to y$&$\mathrm{I_1}$ $(x,y) \to x$&$\mathrm{I_1}$ $(x,y) \to y$\\\hline

 0  & $(x_{n},x_{n-6})$ 100&$(y_{n},y_{n-2})$ 100 &$(x_{n},x_{n-6})$ 100& $(y_{n},y_{n-2},y_{n-12})$ 61\\

    &                      &                      &                     & $(y_{n},y_{n-2},y_{n-13})$ 35\\

 0.5&  $(x_{n},x_{n-6})$ 100&$(y_{n},y_{n-1},y_{n-2})$ 56&$(x_{n},x_{n-6})$ 100& $(y_{n},y_{n-2},y_{n-13})$ 58\\

    &                           &$(y_{n},y_{n-1})$35    &                         & $(x_{n-9},y_{n},y_{n-2},y_{n-13})$ 21\\

 1  &  $(x_{n},x_{n-6})$ 100&$(y_{n},y_{n-1},y_{n-2})$ 56&$(x_{n},x_{n-6})$ 100& $(x_{n-8},y_{n},y_{n-2},y_{n-6})$ 72\\

    &                           &$(y_{n},y_{n-1})$ 44    &                         & $(x_{n},x_{n-8},y_{n},y_{n-2},y_{n-6})$ 17\\

 1.5& $(x_{n},x_{n-6})$ 99&$(y_{n},y_{n-1},y_{n-2})$ 100&$(x_{n},x_{n-6})$ 100& $(x_{n-8},x_{n-12},y_{n},y_{n-2})$ 56\\

   &                      &                           &                       & $(x_{n-8},x_{n-12},y_{n},y_{n-2},y_{n-5})$ 13\\

 2  &  $(x_{n},x_{n-6})$ 100&$(y_{n},y_{n-1},y_{n-2})$ 99 &$(x_{n},x_{n-6})$ 100& $(x_{n},x_{n-8},y_{n},y_{n-2})$ 88\\

 3  &  $(x_{n},x_{n-6})$ 100&$(x_{n},x_{n-9},y_{n},y_{n-2})$ 86&$(x_{n},x_{n-6})$ 100& $(x_{n},x_{n-8},y_{n},y_{n-2})$ 72\\

    &                       &                                   &                         & $(x_{n},y_{n},y_{n-2})$ 27\\

 4  & $(x_{n},x_{n-6})$ 100&$(x_{n},x_{n-8},y_{n},y_{n-2})$ 99 &$(x_{n},x_{n-6})$ 100& $(x_{n-9},y_{n},y_{n-2})$ 90\\\hline\hline

\end{tabular*}
\end{center}
\end{table*}
Again there are no embedding vectors for predicting $x$ that
have components of $y$. We observe that $\mathrm{I_0}$ fails to detect the information
transfer from $x$ to $y$ for small values of $C$ and works well only for $C \ge 3$. On
the other hand, $\mathrm{I_1}$ detects the contribution of $x$ for $C \ge 0.5$.

\begin{figure}[htb]
 \centerline{\includegraphics[height=4.5cm]{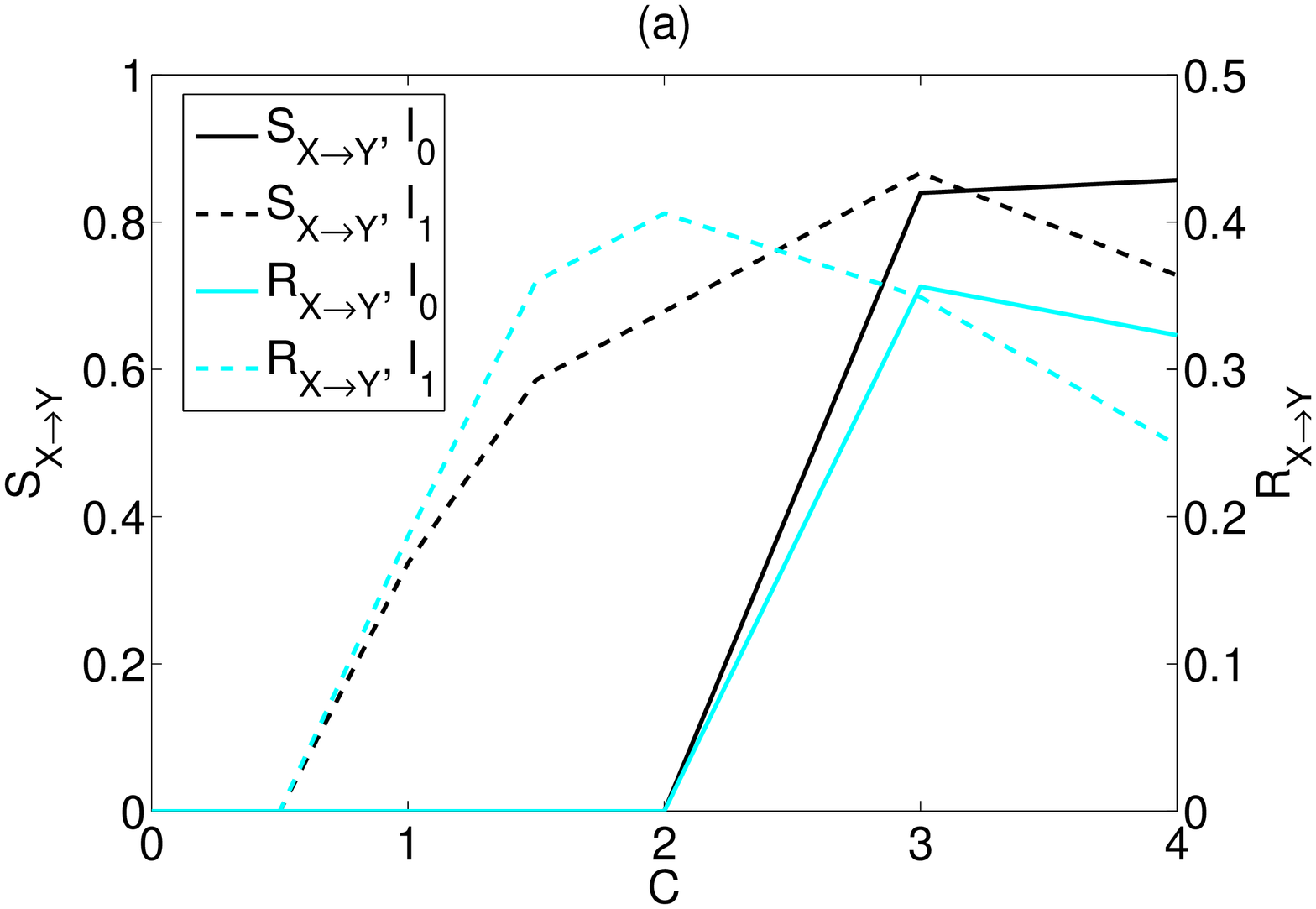}}
\centerline{\includegraphics[height=4.5cm]{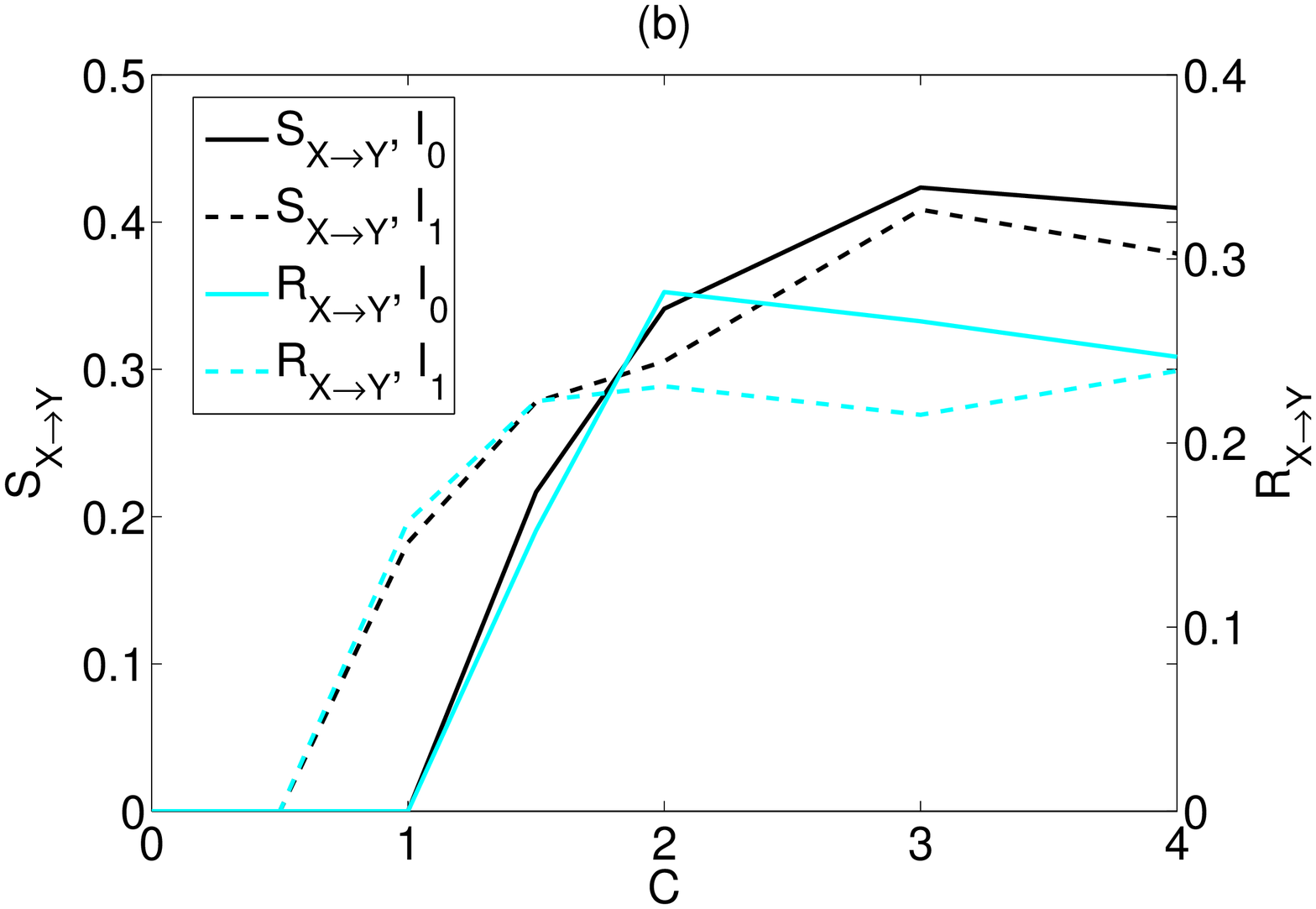}}
 \caption{(Color online) As Fig.~\ref{fig:CHN}(b) and (c) but for the coupled R\"{o}ssler--Lorenz system.}
\label{fig:roslorRS}
\end{figure}

We estimate the average values of $S_{X \rightarrow Y}$ and $R_{X \rightarrow Y}$ on the
100 realizations for the most frequently observed embedding vectors, and the results are
shown in Figure~\ref{fig:roslorRS}(a).
In accordance to the form of $\mathbf{x}_{F}$
we modify the prediction models to make 3-step ahead prediction instead of 1-step.
The contribution of $x$ to the prediction of $y$ is rather high for large
$C$. Again criterion $\mathrm{I_1}$ works better than $\mathrm{I_0}$. $S_{X \rightarrow Y}$ increases
almost monotonically, and only for $C=4$ has a slight decrease. $R_{X \rightarrow Y}$
shows similar behavior, but starts decreasing from $C=3$. For this system, generalized
synchronization occurs for $C \ge 3$ \citep{Quyen99} and this may be the reason for the
decrease of $R_{X \rightarrow Y}$. Under generalized synchronization there exists a
function $\phi$ such that $y=\phi(x)$ ($x$ and $y$ are functionally related) which means
that $\mathbf{x}^1$ and $\mathbf{x}^2$ contain a lot of common information and thus
$I(\mathbf{x}_{F};\mathbf{x}^1 \mid \mathbf{x}^2)$ decreases.

The best embedding dimension for both R\"{o}ssler and Lorenz systems is known to be 3 when no
coupling is present. Thus for the unidirectional coupling from $x$ to $y$ we should
estimate three dimensional vectors comprised only of components of $x$ when predicting
$x$. Both criteria underestimate the embedding dimension in this case (see columns two
and four of Table~\ref{tab:DRroslor}) due to the strictness of the threshold ($A=0.95$) or the
choice of $\mathbf{x}_{F}$. For example, using $A=0.99$ and I$_1$ criterion the embedding
vector contains the component $x_{n-10}$ in addition to the components $x_{n}$ and $x_{n-6}$.

\begin{figure}[htb]
 \centerline{\includegraphics[height=4.5cm]{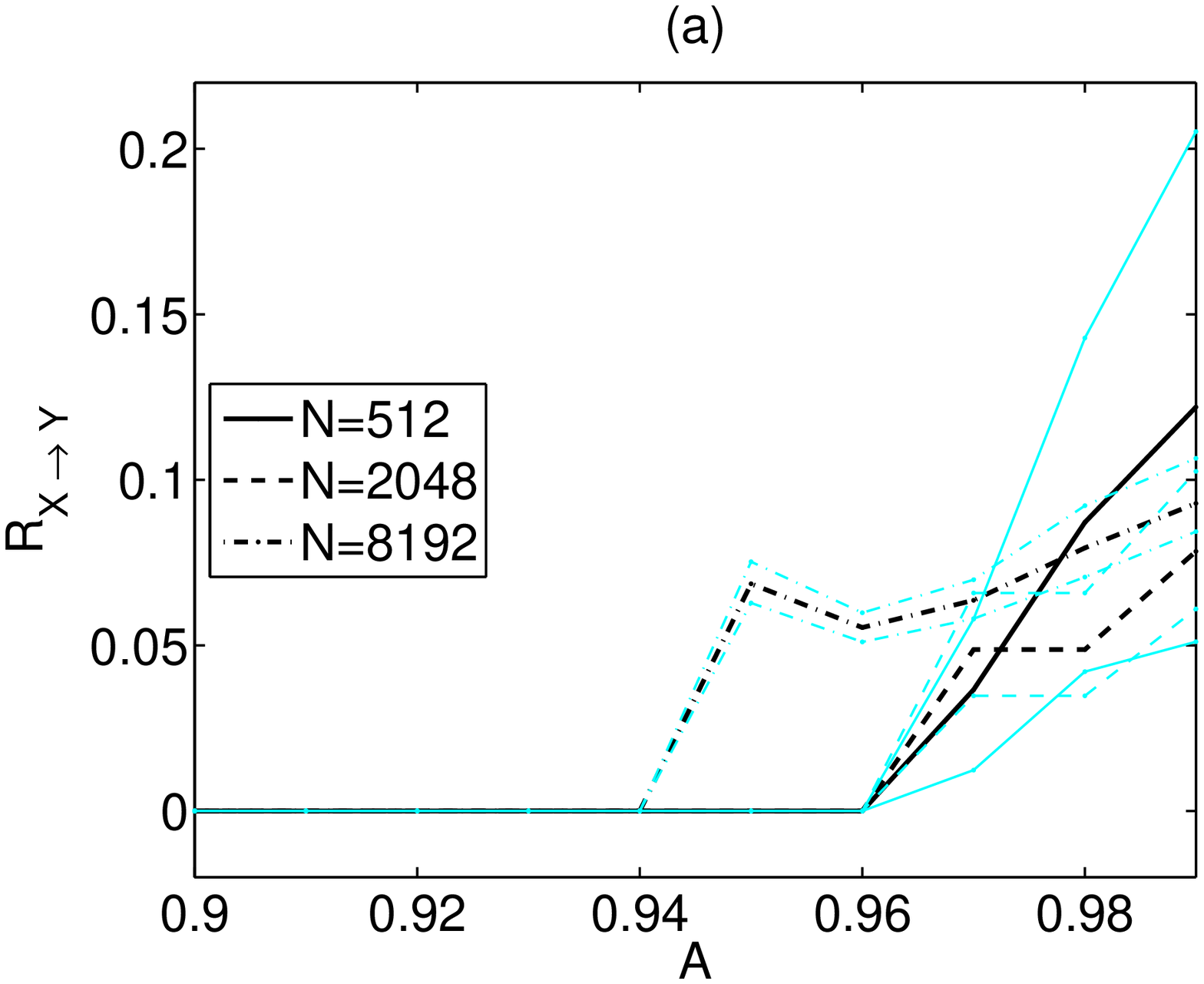}}
\centerline{\includegraphics[height=4.5cm]{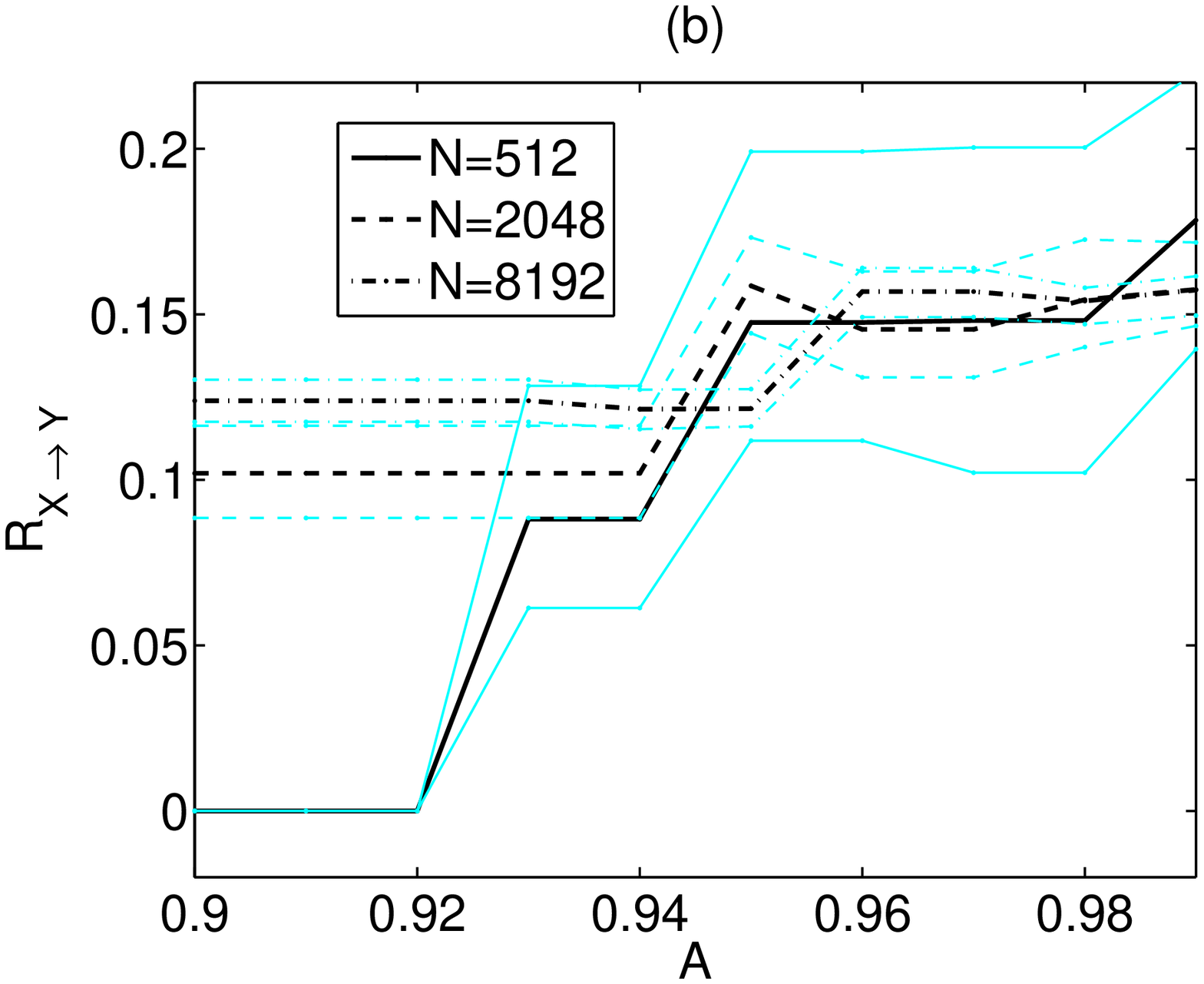} }\caption{(Color online) $R_{X \rightarrow Y}$ vs $A$ for the coupled R\"{o}ssler--Lorenz
system using the $\mathrm{I_1}$ criterion and for different time series length $N$, as shown in the legend. For each $N$, the dark line is the median
and the two gray (cyan on line) lines are the 2.5\% and 97.5\% percentiles. The coupling strength is $C=0.5$ in (a) and $C=1$ in (b).}
\label{fig:roslorRvsA}
\end{figure}

Our simulations have shown that a larger $A$ allows for more iterations of the embedding
scheme and then more components from the driving time series are likely to enter the final
embedding form. For the R\"{o}ssler-Lorenz system, even for very small $C$,
the driving effect is detected by the embedding vector form for larger $A$, and subsequently the directed
coupling measures $S$ and $R$ get positive. In Figure~\ref{fig:roslorRvsA}, we show the
graph of $R_{X \rightarrow Y}$ vs $A$ for the same $\mathbf{x}_{F}$ as above, the I$_1$
criterion, and for $C=0.5$ and $C=1$.
Note that the departure of $R_{X \rightarrow Y}$ from the zero level is succeeded for smaller $A$ as the time series length increases, e.g. for
$C=0.5$ the coupling can be detected ($R_{X \rightarrow Y}$ being positive) for $A\geq0.95$ when $N=8192$ and $A\geq0.97$ when $N=2048$, whereas for
$C=1$ the detection is succeeded for the whole examined range of $A$ and for both $N$. Also, for small time series, as for $N=512$ in
Figure~\ref{fig:roslorRvsA}, the very weak coupling can still be detected but for larger $A$ and with a large variability of the $R_{X \rightarrow
Y}$ measure. For such small time series, a large $A$ does not result in a consistent embedding form as small changes in the data (at each
realization) may give different embedding vectors. This may cause that components from one time series may enter the embedding form even if they do
not really explain the evolution of another time series (the case of uncoupled systems). However, for the R\"{o}ssler-Lorenz system, even for very
large $A$ ($A=0.99$), the embedding scheme did not give rise for spurious causality, i.e. only for $N=512$ the $R$ measure produced positive values
for uncoupled systems and for larger $N$ the $R$ measure was always zero for the uncoupled systems and for the direction of no coupling.

\begin{figure}[htb]
 \centerline{\includegraphics[height=4.5cm]{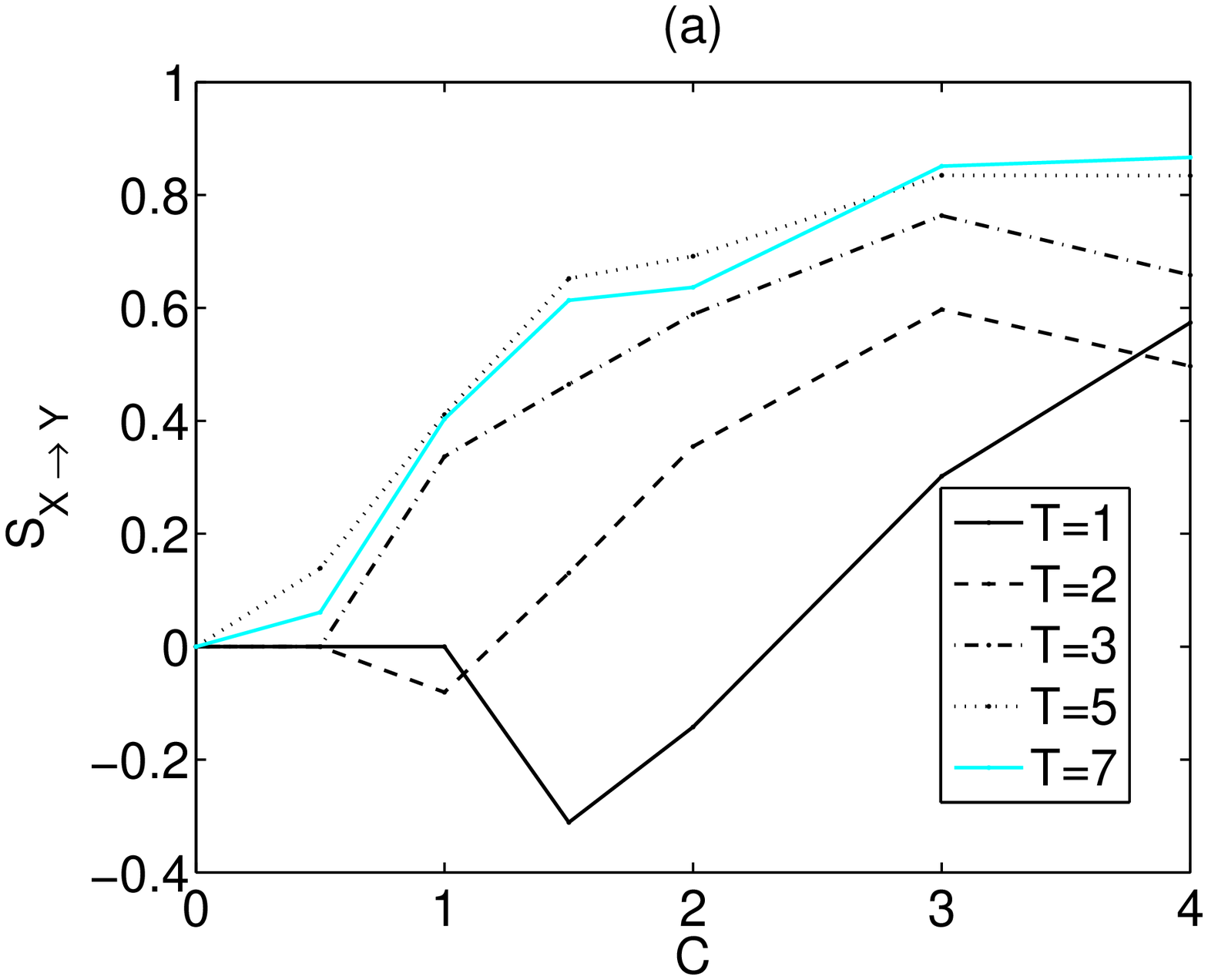}}
\centerline{\includegraphics[height=4.5cm]{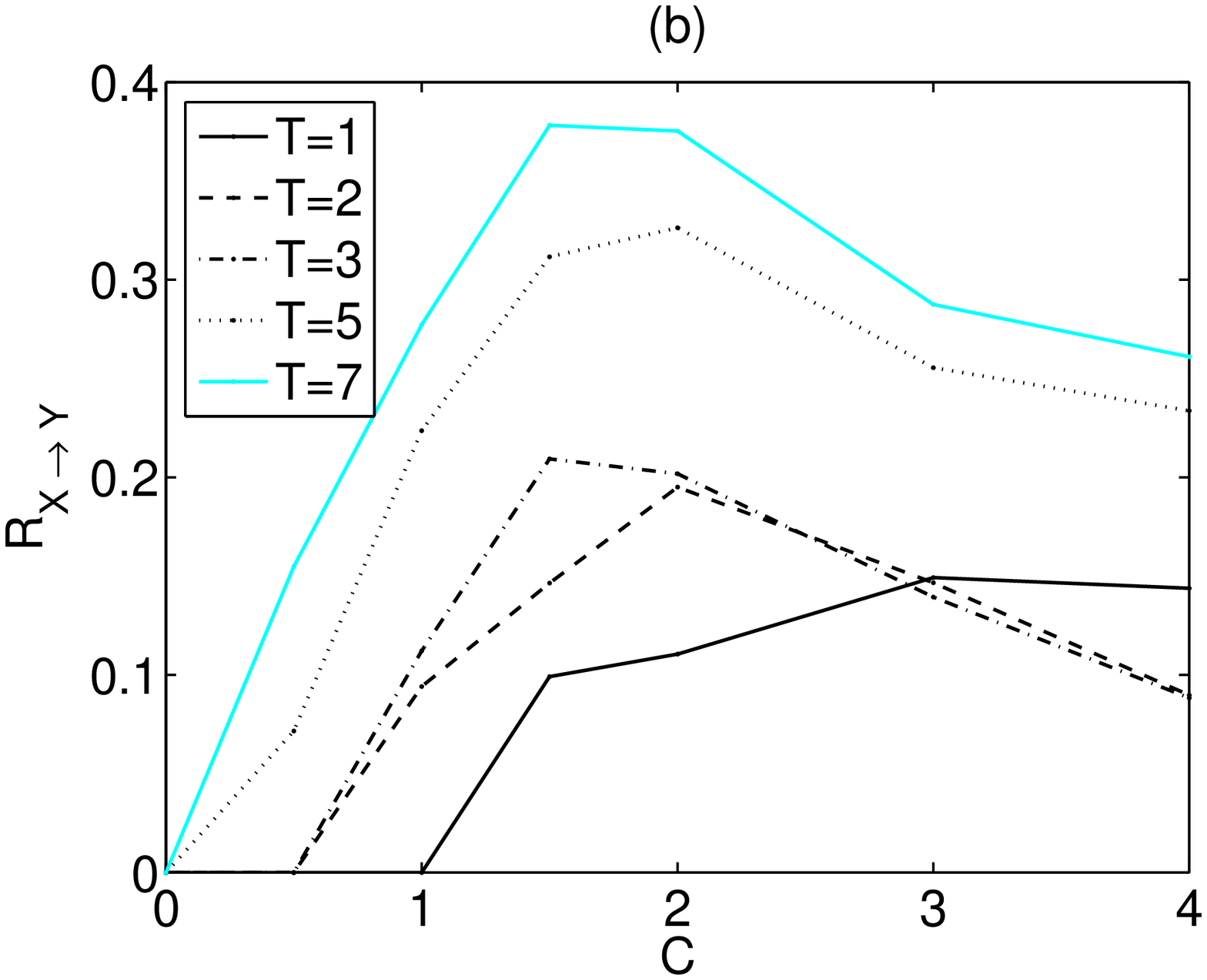}} \caption{(Color online) (a) $S_{X \rightarrow Y}$ vs $C$ for the coupled R\"{o}ssler--Lorenz
system using the $\mathrm{I_1}$ criterion and for different time horizons $T$, as shown in the legend. (b) The same as (a) but for $R_{X \rightarrow
Y}$.} \label{fig:roslorRvsT}
\end{figure}

Regarding the future vector $\mathbf{x}_{F}$, we investigate the effect of the future time horizon $T$ on the embedding vector form. We repeat the
simulations for I$_1$ setting $A=0.95$ and $T=1,\ldots,7$, where the size of $\mathbf{x}_{F}$ increases with $T$. For the uncoupled
R\"{o}ssler-Lorenz system, the embedding vectors do not have mixed components for either direction regardless of $T$, similarly to the embedding
vector forms shown for $T=3$ (Table~\ref{tab:DRroslor}, row for $C=0$). For very weak coupling, the embedding vector forms vary more when $T$ is
larger. The high dimension of the joint vector (due to a large $T$) makes the estimation of entropy terms less accurate, and in the presence of small
coupling, different but close delays are selected as components of the vector form at each run. Besides the differences in the delays of the
components there are always components from the driving system, e.g. components of $x$ are found when explaining $y$ from both $x$ and $y$ in all the
realizations for $C=0.5$ when $T\geq4$. We note also that the number of components of $x$ and $y$ in the embedding vector at each $C$ are the same
for $T\geq4$. Moreover, the driving effect can be better detected when $T$ is larger, as shown for the $S_{X \rightarrow Y}$ and $R_{X \rightarrow
Y}$ measures in Figure~\ref{fig:roslorRvsT}.
For $T=1$, $R_{X \rightarrow Y}$ finds no driving effect $X \rightarrow Y$ unless $C\geq1.5$, whereas for $T=2,3$
it detects driving effect also for $C=1$ and for $T\geq4$ also for $C=0.5$. The measure $S_{X \rightarrow Y}$ behaves similarly but gives
negative values instead of positive when $T=1,2$ and $C$ is small, which means that the prediction
is actually better when subtracting the $x$ components from the embedding vector form. Thus though there
is coupling for $C=1,1.5$, the use of $y$ components alone is sufficient to predict $y$ a couple of steps
ahead ($T=1,2$). This is a drawback of the local prediction method in $S$ that uses only the closest
neighboring point.

Next we investigate the effect of noise on the embedding schemes.
In Table \ref{tab:DRroslornoise}, we show the embedding vector forms and their frequencies for
the embedding schemes I$_0$ and I$_1$ when 20\% observational noise is added to each variable,
fixing the other parameters ($A=0.95$, $T=3$, $N=4096$).
\begin{table*}[htb]
\caption{\label{tab:DRroslornoise}Embedding vectors and their frequency of occurrence for the unidirectionally coupled R\"{o}ssler--Lorenz systems with
noise.}
\begin{center}
\begin{tabular*}{1\textwidth}{@{\extracolsep{\fill}}lrrrr}
\hline\hline$C$&$\mathrm{I_0}$ $(x,y) \to x$&$\mathrm{I_0}$ $(x,y) \to y$&$\mathrm{I_1}$ $(x,y) \to x$&$\mathrm{I_1}$ $(x,y) \to y$\\\hline

 0  & $(x_{n},x_{n-6},x_{n-9})$ 77&$(y_{n},y_{n-1},y_{n-3},y_{n-7})$ 22&$(x_{n},x_{n-6},x_{n-9})$ 69& $(y_{n},y_{n-1},y_{n-3},y_{n-13})$ 17\\

    & $(x_{n},x_{n-6},x_{n-8})$ 7&$(y_{n},y_{n-1},y_{n-2},y_{n-6})$ 12    &$(x_{n},x_{n-6},x_{n-10})$ 15& $(y_{n},y_{n-1},y_{n-3},y_{n-10},y_{n-13})$ 15\\

 0.5&  $(x_{n},x_{n-6},x_{n-9})$ 81&$(y_{n},y_{n-2},y_{n-4})$ 34&$(x_{n},x_{n-6},x_{n-9})$ 81& $(y_{n},y_{n-1},y_{n-3},y_{n-7})$ 36\\

    &                                    &$(y_{n},y_{n-2},y_{n-5})$ 23    &                                  & $(y_{n},y_{n-2},y_{n-5})$ 16\\

 1  &  $(x_{n},x_{n-6},x_{n-9})$ 86&$(y_{n},y_{n-2},y_{n-4})$ 42&$(x_{n},x_{n-6},x_{n-9})$ 82& $(x_{n-8},x_{n-12},y_{n},y_{n-2},y_{n-4})$ 12\\

    &                                    &$(y_{n},y_{n-2},y_{n-3})$ 31    &                                  & $(x_{n-8},x_{n-12},y_{n},y_{n-2},y_{n-5})$ 9\\

 1.5& $(x_{n},x_{n-6},x_{n-9})$ 81&$(x_{n-8},y_{n},y_{n-2},y_{n-4})$ 46&$(x_{n},x_{n-6},x_{n-9})$ 79& $(x_{n-8},x_{n-12},y_{n},y_{n-2},y_{n-5})$ 27\\

    &                                   &$(x_{n-8},y_{n},y_{n-2},y_{n-3})$ 24&$(x_{n},x_{n-6},x_{n-10})$ 10& $(x_{n-8},x_{n-13},y_{n},y_{n-2},y_{n-5})$ 17\\

 2  &  $(x_{n},x_{n-6},x_{n-9})$ 78&$(x_{n-8},x_{n-12},y_{n},y_{n-2})$ 20&$(x_{n},x_{n-6},x_{n-9})$ 78& $(x_{n-8},x_{n-12},y_{n},y_{n-2},y_{n-7})$ 11\\

    &  $(x_{n},x_{n-6},x_{n-8})$ 8&$(x_{n-8},x_{n-11},y_{n},y_{n-2})$ 14 &$(x_{n},x_{n-6},x_{n-10})$ 10& $(x_{n},x_{n-7},y_{n},y_{n-2},y_{n-5})$ 11\\

 3  &  $(x_{n},x_{n-6},x_{n-9})$ 69&$(x_{n},x_{n-4},y_{n},y_{n-2})$ 42&$(x_{n},x_{n-6},y_{n-5})$ 78& $(x_{n},x_{n-4},y_{n},y_{n-2},y_{n-14})$ 17\\

    &  $(x_{n},x_{n-6},x_{n-8})$ 7&$(x_{n},x_{n-5},y_{n},y_{n-2})$ 20    &$(x_{n},x_{n-6},x_{n-9})$ 8& $(x_{n},x_{n-5},y_{n},y_{n-2})$ 13\\

 4  & $(x_{n},x_{n-6},x_{n-9})$ 69&$(x_{n},x_{n-6},y_{n},y_{n-2})$ 21&$(x_{n},x_{n-6},x_{n-9})$ 29& $(x_{n-9},x_{n-13},y_{n},y_{n-2})$ 20\\

    & $(x_{n},x_{n-6},x_{n-10})$ 7&$(x_{n-2},x_{n-9},y_{n},y_{n-2})$ 11    &$(x_{n},x_{n-6},y_{n-6},y_{n-8})$ 14& $(x_{n},x_{n-7},y_{n},y_{n-2},y_{n-8})$ 10\\\hline\hline

\end{tabular*}
\end{center}
\end{table*}
The presence of noise changes the results on the noise-free system. More forms of
embedding vectors are observed mostly because a component is replaced by another
temporally close component. For example, component $x_{n-9}$ is often replaced by
$x_{n-8}$ and $x_{n-10}$. Another interesting result is that criterion $\mathrm{I_1}$ for
$C=3$ chooses the component $y_{n-5}$ for explaining $x$. Since the coupling is
unidirectional this seems wrong. However, the two variables are functionally related for
this coupling strength, and the inclusion of $y_{n-5}$ may be explained by a better
noise-reduction. For $y$, both criteria give a large variety of embedding vector forms.
In fact for some cases the most observed embedding vector has a frequency near 10\%.
Criterion $\mathrm{I_0}$ is a bit more consistent but again fails to detect the coupling
for $C=1$. The $S_{X \rightarrow Y}$ and $R_{X \rightarrow Y}$ measures give the same
pattern as for the noise-free case but take smaller values (see
Figure~\ref{fig:roslorRS}(b)).

For the peculiar case
of predicting $x$ for $C=3$ using the embedding scheme $\mathrm{I_1}$, the value of $S_{Y
\rightarrow X}$ is 0.25 (not shown), which indicates significant coupling
from $y$ to $x$. This is not very worrying, because most information transfer measures
tend to find such erroneous couplings when there is generalized synchronization.

\subsection{Unidirectionally coupled Mackey-Glass}

The last deterministic system in our simulation study is that of unidirectionally coupled identical Mackey-Glass equations \cite{Senthilkumar08} given by
\begin{equation*}
\label{eqap:couplMackeyGlass}
\begin{array}{lll}
\dot x(t)&=&{0.2x(t-\Delta_1) \over 1+x(t-\Delta_1)^{10}}-0.1x(t)\\
\dot y(t)&=&{0.2y(t-\Delta_2) \over 1+y(t-\Delta_2)^{10}}-0.1y(t)+C{x(t-\Delta_1) \over 1+x(t-\Delta_1)^{10}}.
\end{array}
\end{equation*}
For $\Delta_1, \Delta_2=$ 17, 30 and 100, coupling strength $C=[0, 0.05, 0.1, 0.15, 0.2, 0.3, 0.4,
0.5]$ and all possible combinations of driving-driven time series we study the embedding
vector forms. Because the complexity of the individual uncoupled system differs significantly
with the value of $\Delta$ (meaning $\Delta_1$ or $\Delta_2$),
the choice of $\mathbf{x}_{F}$ is of great importance, and we tried different
$\mathbf{x}_{F}$ with regard to both the coupling strength and the systems under
investigation.

At first, a limited study of few realizations was performed for three choices of
$\mathbf{x}_{F}$,
\begin{enumerate}
    \item $\mathbf{x}_{F}=(y_{n+1}),$
    \item $\mathbf{x}_{F}=(y_{n+1},y_{n+2},\ldots,y_{n+\tau_1}),$
    \item $\mathbf{x}_{F}=(y_{n+1},y_{n+2},\ldots,y_{n+\tau_2}),$
\end{enumerate}
where $\tau_1$ is the lag that gives the first minimum of mutual information and $\tau_2$
the first maximum. These values were estimated for each coupling strength separately and
we summarily present our conclusions. We observed that on the uncoupled systems where we
have indications of the proper embedding dimension from the literature, the first choice
of $\mathbf{x}_{F}$ seriously underestimated it, especially for $\Delta=100$ where the
embedding dimension was estimated as 2, when an appropriate value should be greater than
the fractal dimension of about 7. Also, there were a lot of cases where despite the
presence of moderate and strong coupling the embedding vectors for the driven system
($y$) did not have components from the driving one ($x$). This was also observed for
the R\"{o}ssler-Lorenz system (see Figure~\ref{fig:roslorRvsT} for $T=1$).
The second choice gave
reasonable embedding dimensions on the uncoupled systems, 3 for $\Delta=17$, 4 for
$\Delta=30$ and 9 for $\Delta=100$, while the third choice gave respectively 3, 5 and 10.
The values of both these choices are in order, being always just greater than the
respective fractal dimensions. For the uncoupled time series of $\Delta=100$ the value of
$\tau_1$ was 14 and for $\tau_2$ it was 26 and both increased slightly for moderate
coupling. These values are quite large and the estimation of mutual information on such
multidimensional vectors including all the lags from 0 to $\tau_1$ or $\tau_2$ is tricky
and computationally demanding. A nice and simple alternative that we opted to use
ultimately was $\mathbf{x}_{F}=(y_{n+1},y_{n+\tau_1},y_{n+\tau_2})$. In this way, we
include information further into the future and restrict $\mathbf{x}_{F}$ to a three
dimensional vector, which is efficient for mutual information estimation.

\begin{figure*}[htb]
 \centerline{\includegraphics[height=9cm]{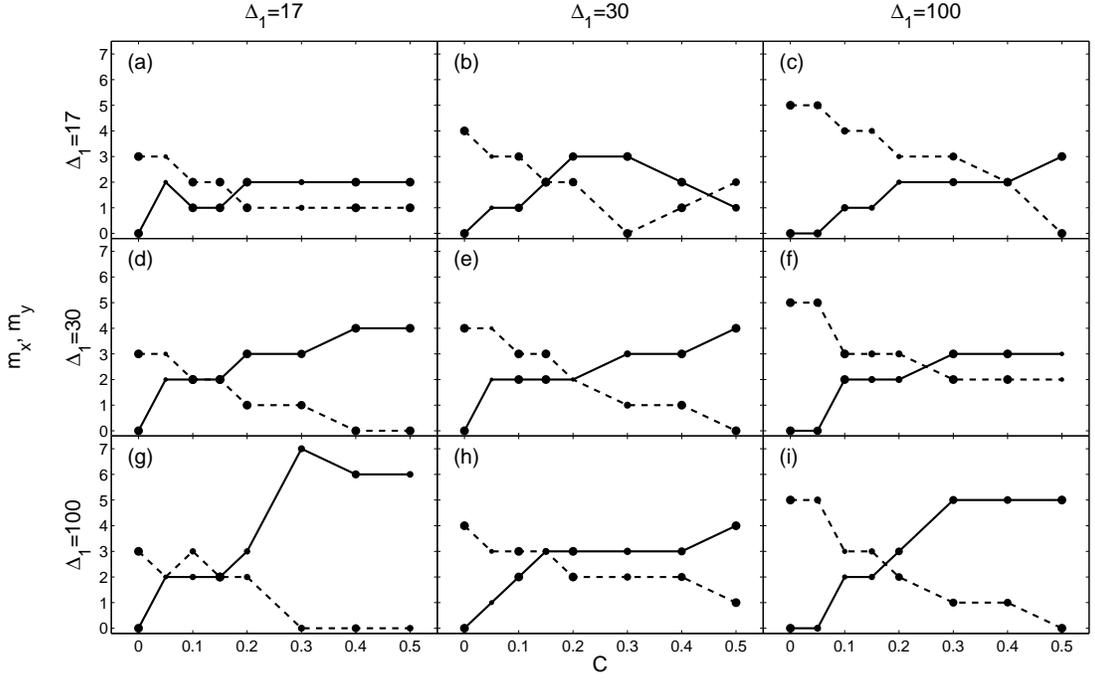}}
\caption{Number of components from each time series for mixed embedding vectors with
regard to the coupling strength for the Mackey-Glass systems. Solid line denotes the
driving system and dashed line the driven system. The panels are ordered for the systems
indicated by $\Delta_1$ for the driving system and $\Delta_2$ for the driven system. The
size of the markers indicates the frequency of occurrence of the most frequently selected
embedding vectors (same size of the markers for each $C$ value). For example, the size of
the marker in panel (e) for $C=0$ regards the largest frequency 100, whereas in the same
panel for $C=0.2$ is the smallest frequency in these results equal to 42.}
\label{MGDIFDELTATAU}
\end{figure*}

Using this form of $\mathbf{x}_{F}$ we performed 100 realizations for cross-modelling of
$x$ from $x$ and $y$, and $y$ from $x$ and $y$. The time series length was set to $N$=4096, $L_{x}
= L_{y}=50$ and the $\mathrm{I}_1$ criterion was used. All combinations of $\Delta_1$,
$\Delta_2$ and both coupling directions were examined and we marked the number of components from each
system that were selected in the embedding vectors. The results for explaining $y$ from
$x$ and $y$ are shown in Fig.~\ref{MGDIFDELTATAU}, where each line regards the number of
components from each system that are used in forming the embedding vector.
We see that
for all combinations of $\Delta_1$, $\Delta_2$, as coupling increases the number of
components of the driven system reduces and this of the driving system increases, as we
would expect. The embedding dimension for $\Delta_2=100$ is again underestimated, being 5
for the uncoupled case, but the conclusions regarding coupling direction and strength are
sound.

For the inverse direction, i.e. cross-modelling of $x$ from $x$ and $y$, the embedding vectors were comprised almost exclusively from components of
$x$. One exception was for $\Delta_1=17$, $\Delta_2=30$ and $C=0.5$, where coupling was detected in the wrong direction, from $y$ to $x$. For such
strong coupling there is generalized synchronization between the two systems and components from the two systems can be interchanged.

The $S$ and $R$ measures gave qualitatively the same results as these in Figure~\ref{MGDIFDELTATAU}
 and therefore they are not presented.

\subsection{Coupled linear stochastic systems}

The proposed embedding scheme is based on entropies and makes no assumption for deterministic
or nonlinear underlying dynamics, so in principle it can also be used in stochastic systems.
Popular models for stochastic multivariate time series are the vector autoregressive (VAR) models
of order $m$, assigning equal number of delays to all $p$ variables, and the dynamic regression
(DR) or distributed lag models that allow for different delays for each variable. Commonly the
VAR and DR models are best known in their linear form \cite{Wei06,Pankratz91}.

We test the proposed embedding scheme in identifying the correct components present in the
form of a linear DR model of order one in both variables
\begin{equation*}
\begin{array}{lll}
x_{n+1} & = & 0.5x_{n} + \epsilon_{1,n} \\
y_{n+1} & = & 0.5y_n + C x_n + \epsilon_{2,n}
\end{array}
\end{equation*}
Only the second variable $y$ depends on $x$ and the coefficient $C$ determines the strength
of dependence.
The random components $\epsilon_{1,n}$ and $\epsilon_{2,n}$ are independent to $x$ and $y$ and to
themselves for any lag and have mean zero and standard deviation one.
The embedding vector forms and their frequencies for the embedding scheme I$_1$ explaining
$x$ from $x$ and $y$, and $y$ from $x$ and $y$, are shown in Table~\ref{tab:VAR} for $C=0.0,0.1,0.2,0.3,0.4,0.5$
and two time series lengths $N=512$ and $N=4096$ ($A=0.95$, $T=1$).
\begin{table*}[htb]
\caption{\label{tab:VAR}
Embedding vectors and their frequency of occurrence for the VAR system in two variables of order one
for $N=512$ and $N=4096$.}
\begin{center}
\begin{tabular*}{1\textwidth}{@{\extracolsep{\fill}}lrrrr}
\hline\hline $C$ & \multicolumn{2}{c}{$\mathrm{I_1}$ $(x,y) \to x$} & \multicolumn{2}{c}{$\mathrm{I_1}$ $(x,y) \to y$} \\
& $N=512$ & $N=4096$ & $N=512$ & $N=4096$ \\\hline
 0 & $(x_{n})$ 9 & $(x_{n})$ 46 & $(y_{n})$ 6 & $(y_{n})$ 44 \\
& $(x_{n},y_{n-1})$ 3 & $(x_{n},y_{n-3})$ 6 & $(x_{n-1},y_{n})$ 3 & $(x_{n-2},y_{n})$ 7 \\\hline

 0.1  & $(x_{n})$ 8 & $(x_{n})$ 42 & $(x_{n},y_{n})$ 6 & $(y_{n})$ 33 \\
& $(x_{n},y_{n-1})$ 4 & $(x_{n},y_{n})$ 7 & $(x_{n},y_{n},y_{n-3})$ 4 & $(x_{n},y_{n})$ 20 \\\hline

 0.2  & $(x_{n})$ 12 & $(x_{n})$ 44 & $(x_{n},y_{n})$ 13 & $(x_{n},y_{n})$ 83 \\
& $(x_{n},x_{n-4})$ 4 & $(x_{n},y_{n})$ 6 & $(x_{n},y_{n},y_{n-2})$ 4 & $(x_{n},y_{n},y_{n-5})$ 3 \\\hline

 0.3  & $(x_{n})$ 7 & $(x_{n})$ 46 & $(x_{n},y_{n})$ 27 & $(x_{n},y_{n})$ 95 \\
& $(x_{n},x_{n-4})$ 4 & $(x_{n},y_{n})$ 5 & $(x_{n},x_{n-2},y_{n})$ 8 & $(x_{n},y_{n},y_{n-5})$ 1 \\\hline

 0.4  & $(x_{n},x_{n-4})$ 4 & $(x_{n})$ 46 & $(x_{n},y_{n})$ 53 & $(x_{n},y_{n})$ 100 \\
& $(x_{n})$ 3 & $(x_{n},y_{n-3})$ 6 & $(x_{n},x_{n-2},y_{n})$ 6 &  \\\hline

 0.5  & $(x_{n})$ 7 & $(x_{n})$ 46 & $(x_{n},y_{n})$ 74 & $(x_{n},y_{n})$ 100 \\
& $(x_{n},x_{n-4})$ 4 & $(x_{n},y_{n-1})$ 7 & $(x_{n},x_{n-2},y_{n})$ 6 &  \\\hline\hline

\end{tabular*}
\end{center}
\end{table*}
For the direction of no driving effect, as well as for $C=0$, the embedding scheme
gives varying embedding vector forms when $N=512$. When $N=4096$, it identifies
correctly the sole regressor in about half of the realizations, whereas for the rest realizations
the model order is overestimated adding another component in the embedding vector from $x$ or $y$.
Overestimation of the model order is often observed by classical model order criteria, such as
AIC and BIC, but for shorter time series \cite{Wei06}, indicating that the proposed embedding scheme
is more data demanding than other standard statistical techniques for model order selection.
For any of the embedding vector forms, the absence of true driving effect is correctly identified
by the $S$ and $R$ measures being zero. In the direction of driving effect, the correct embedding
vector $(x_{n},y_{n})$ is found with a frequency that increases with $C$, slowly for $N=512$ and
rapidly for $N=4096$. The measures $S$ and $R$ increase accordingly and the mean $S$ and $R$
get positive for $C\geq0.2$.

The coupling is detected similarly with the embedding scheme for a DR model of order one in the
first variable and order two in the second variable
\begin{equation*}
\begin{array}{lll}
x_{n+1} & = & 0.3 x_{n} - 0.5 x_{n-1} + \epsilon_{1,n} \\
y_{n+1} & = & 0.4 x_n - 0.3 y_{n-1} + \epsilon_{2,n}
\end{array}
\end{equation*}
where $\epsilon_{1,n}$ and $\epsilon_{2,n}$ are independent as above. The embedding vector
for explaining $x_{n+1}$ is correctly identified as $(x_{n},x_{n-1})$ in 12 out of 100 realizations
when $N=512$ and increases linearly with $N$ (25 for $N=1204$, 50 for $N=2048$ and 78 for
$N=4096$), where again the embedding vectors are augmented in the rest realizations.
For explaining $y_{n+1}$ the correct embedding vector $(x_{n},y_{n-1})$ is found in 14 realizations
when $N=512$ and increases again linearly with $N$. Thus the proposed embedding scheme, and
subsequently the $S$ and $R$ measures, can detect the correct direction of coupling but require
considerable data size.

\subsection{Significance of $S$ and $R$ measures}
Reliable detection of directed coupling requires that the coupling measure is examined for
its significance. For example, a positive $R_{X \rightarrow Y}$ can be considered to indicate
a driving effect of $X$ on $Y$ only if it exceeds a threshold of significance and, on the other
hand, it should not exceed it if the driving effect is not present.
In the lack of analytic null probability distribution of the coupling measure, i.e., distribution
under the null hypothesis of no coupling,
surrogate techniques have been proposed \cite{Thiel06,Faes08}.

We use the simple technique of time shifted surrogates in \cite{Faes08}, and for
each randomly selected time index $i$ (requiring $i>100$) the surrogate couple of time
series is as the original but the first time series is time shifted to
$(x_i,x_{i+1},\ldots,x_{N},x_1,x_2,\ldots,x_{i-1})$. The embedding scheme on the surrogate
bivariate time series when explaining $y$ from $x$ and $y$ contains components from $y$,
and components from $x$ enter the embedding vector by chance and regardless of the coupling strength.
Thus the measures $S_{X \rightarrow Y}$ and $R_{X \rightarrow Y}$ computed on surrogate bivariate
time series are predominately zero and may get a positive value whenever a component from $x$ enters
the embedding scheme. Obviously the null distribution of $S_{X \rightarrow Y}$ and $R_{X \rightarrow Y}$
is not symmetric and the decision for the surrogate data test should be taken on the basis of rank
ordering rather than $z$-score.

\begin{figure}[htb]
 \centerline{\includegraphics[height=4cm]{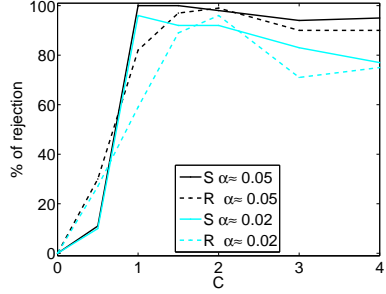}}
\caption{(Color online) Percentage of rejection vs coupling strength $C$ for the surrogate data test of no causality from $x$ to $y$ for the coupled
R\"{o}ssler-Lorenz system ($A=0.95$, $T=3$, $N=4096$). The measures $S$ and $R$ used as test statistics and the significance levels $\alpha$ are given
in the legend.} \label{fig:rossur}
\end{figure}

We illustrate the correct significance and good power of the measures $S_{X \rightarrow Y}$ and $R_{X \rightarrow Y}$
from criterion I$_1$ on the coupled R\"{o}ssler-Lorenz system ($A=0.95$, $T=3$, $N=4096$).
In Figure~\ref{fig:rossur}, we show the estimated probability
(relative frequency from 100 realizations) of rejection of the null hypothesis of no coupling
using rank ordering.
The test is one-sided and if, say, the
$S_{X \rightarrow Y}$ value computed on the original pair of time series is larger than all
$S_{X \rightarrow Y}$ values computed on $M=40$ surrogate pairs, the null hypothesis is rejected at the
significance level $\alpha$ of about 0.02 (the $p$-value of the test at this case is
$1-(41-0.326)/(41+0.348)=0.0163$, using the corrected rank estimation in \cite{Yu01}).
In Figure~\ref{fig:rossur} the results are shown also for $\alpha \approx 0.05$
(when the statistic for the original data set is second largest we have $p=0.0405$).
The graphs of the probability of rejection of no coupling vs coupling strength $C$ are in agreement
with the results on $S_{X \rightarrow Y}$ and $R_{X \rightarrow Y}$ in Figure~\ref{fig:roslorRS},
and show that though $S_{X \rightarrow Y}$ and $R_{X \rightarrow Y}$ do not reach maximum values
for $C=1$ the driving is detected with the same great confidence as for larger $C$. For the time series
length $N=4096$, the power of the surrogate data test is small when $C=0.5$ as
even the largest probability of rejection found for $R_{X \rightarrow Y}$ is at 0.3.
Higher statistical confidence of rejection can be achieved when $N$ is doubled in agreement
with positive $R_{X \rightarrow Y}$ in Figure~\ref{fig:roslorRvsA}a.

\section{Application to EEG}\label{ivlach:sec5}

\begin{figure}[htb]
 \centerline{\includegraphics[height=7cm]{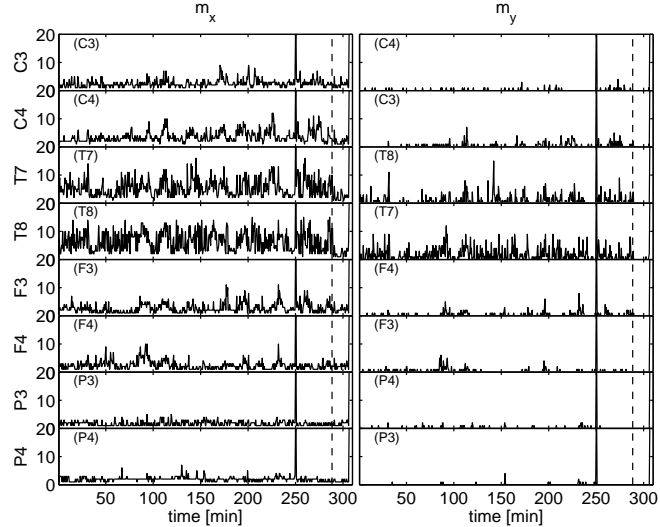}}\caption{Number of components from each channel for mixed embedding vectors for patient
A. Each row of 2 panels corresponds to a mixed embedding for explaining the signal of the
channel indicated by the label on the left of the panels. Each left panel shows the
number of components in the mixed embedding that are derived from the channel being
explained and the right panel those from its antidiametrical. For example, the upper left
panel has the number of components of C3 and the upper right those of C4 used in the
embedding explaining C3, both as a function of time. The dashed vertical line shows the
seizure onset.}\label{EEGGTKLAP}
\end{figure}

\begin{figure}[htb]
 \centerline{\includegraphics[height=7cm]{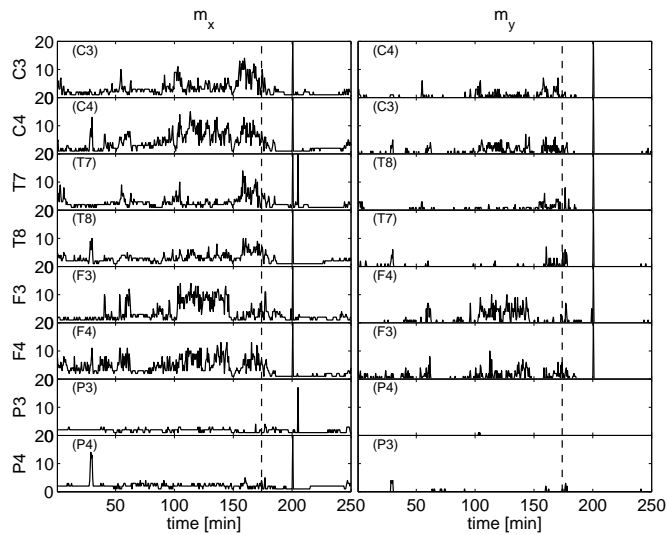}}
\caption{The same as Fig.~\ref{EEGGTKLAP} but for patient B.} \label{EEGNILLAP}
\end{figure}

We apply the non-uniform multivariate embedding procedure on EEG recordings of two
epileptic patients, one with generalized tonic clonic seizure and one with partial complex seizure.
The analysis was restricted to specific EEG channels and the goal was to study
information transfer between anti-diametrical areas of the brain in the left and right
hemisphere. Four channels are selected on each hemisphere and for each channel
the average of its 4 nearest channels is subtracted. This transformation is called surface Laplacian
estimation and it is considered to improve the spatial resolution of scalp
potentials by reducing the common activities between neighboring electrodes, refining the data
and removing possible artifacts \citep{Hjorth75}. The channels used are
C3, T7, F3, and P3 on the left hemisphere and the corresponding C4, T8, F4 and P4 on the
right hemisphere. The sampling time is 0.01 sec and the record lengths are approximately 5 hours for
patient A and 4 hours for patient B. Taking the channels in pairs (each one with its
antidiametrical) we split the time series into segments of length $N=3000$ data points
(corresponding to 30 seconds duration) and apply the embedding scheme for mixed
modelling. The threshold is $A=0.95$, the maximum lags are chosen $L=20$ and
$\mathbf{x}_{F}=x_{n+1}$ (this selection is justified by the vanishing of delayed MI after
the first lags). For each left-right pair of channels we obtain embedding
vectors for explaining each of the two channels. The number of components from each
channel (left, right) in the mixed embedding vector for all 4 channel pairs and the two
directions are given in Fig.~\ref{EEGGTKLAP} for patient A and Fig.~\ref{EEGNILLAP} for
patient B.


There are some common conclusions to be drawn for both patients. There seems to be little
or no information exchange between channels P3 and P4 as indicated by the embedding
vectors having almost always no components from the antidiametrical channel (bottom
panels in Fig.~\ref{EEGGTKLAP} and Fig.~\ref{EEGNILLAP}).
For channel C4, the embedding vectors tend to have more components from
both C3 and C4, than for channel C3 (more clearly for patient A), indicating possibly
difference in the activity of these areas of the brain. A noteworthy observation is that
soon after seizure onset there is no information exchange (or at least not strong enough
to be detected by our approach) between the two hemispheres, and in all cases the mixed
embedding vectors have no components from the antidiametrical channel (bottom parts on
all panels on the right of the dashed line of seizure onset). This is more clearly
visible in patient B where the record extends further after the seizure. This finding
is in agreement with recent reports on resetting of the brain activity at the seizure end
\cite{Iasemidis04,Schindler07}.

\begin{figure}[htb]
 \centerline{\includegraphics[height=6.5cm]{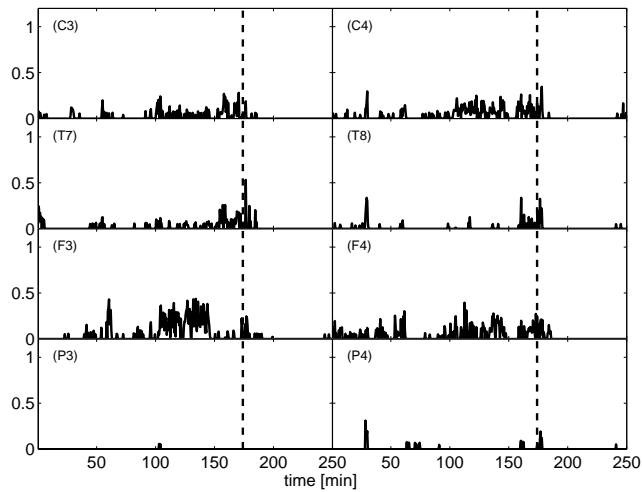}}
 \caption{The $R$ measure for the mixed embeddings of Fig.~\ref{EEGNILLAP} for patient B. $R_{C4\rightarrow C3}$ is shown in top left panel and
$R_{C3\rightarrow C4}$ in top right panel. Similarly for the other panels, as indicated by the inset text.} \label{EEGNILLRmea}
\end{figure}

For patient A there is diversity in the embedding vectors with regard to the channels.
The dimensions for channels C3 and P3 as well as C4 and P4 are similar in form, while for
T7 and T8 there is a substantially higher number of components, again indicating
difference in the activity of these areas of the brain. There does not seem to be any
change in the embedding vector forms as the seizure onset approaches. In patient B we
see that after seizure onset the dimension of the embedding vectors decreases
dramatically, while right before the onset there is a slight increase in the number of
components from both channels contributing to the embedding that explains channels C3, T7
and C4, T8 (Fig.~\ref{EEGNILLAP}, top 4 panels). For channel F3, there is a large increase
in the vector dimension in the time interval 70--25 minutes before seizure and for
channel F4 the embedding vectors have consistently, more or less, high dimensions.
The respective profiles of $S$ and $R$ measures are in complete agreement with the profiles
of the number of components of the mixed embedding, as shown for patient B in
Figure~\ref{EEGNILLRmea} (similarly for patient A, not shown).

The difference in the results of corresponding channels between the two patients may
be due to different seizure types. Patient A has
generalized tonic clonic seizure and we expect the results from all channels to be more or less the same, while patient B has partial complex seizure
and we expect the results to differ from channel to channel.

\section{Discussion}
\label{ivlach:sec6}

The proposed method for reconstruction with a purpose works well in the tested systems.
Our embedding scheme
uses information measures in order to decide at each step which component to be included
in the embedding vector. The estimation of mutual information on high-dimensional vector
variables introduces large bias and variance. We used the estimator of $K$-nearest neighbor
distances that is rather stable for high dimensions. Further, we found that the use of the
conditional mutual information instead of the mutual information reduces the bias.
The embedding scheme using the conditional mutual information turned out to be more
efficient in selecting the embedding vector components. This was demonstrated in simulated
chaotic systems of different complexity.

The termination of the embedding procedure is done using a threshold criterion and we
found that a strict threshold, such as $A=0.95$, may lead to under-embedding for continuous systems.
We note however that for the application of information transfer detection a strict
threshold is needed. If one uses a larger threshold value than 0.95, this would lead to
quite large embedding dimensions and many components that contribute very little would be
included. We found that this does not necessarily impair the accurate detection of coupling,
but under specific data conditions, e.g. noisy data, it may lead to spurious coupling detection.

The strict threshold and the progressive nature of our method guarantees that from all
the components of the embedding vector, only few (if any) have small contribution, and
this is what allows the clear detection of information transfer. If for example we have two
time series from a coupled system with minimum embedding dimension $m_1$ and $m_2$,
respectively, and the embedding vectors for cross-prediction are large enough to contain
$m_1$ components from the first and $m_2$ from the second, then the sub-embedding used in
the estimation of measures $S$ and $R$ would theoretically be equally good to the
complete embedding, and the measures would not be able to clearly detect the coupling.
The embedding that we propose here is of a minimum ``adequate'' dimension in the sense
that it is sufficient to correctly detect interdependencies and free of redundant
information that may confuse the applied measures.

The proposed embedding scheme can be used for different purposes in univariate and
multivariate time series analysis, such as invariant estimation, cross-prediction
and mixed prediction. In particular, it is capable of detecting the coupling and
measuring its strength, as demonstrated on different nonlinear deterministic systems,
as well as stochastic linear systems,  and on epileptic multi-channel EEG.
For the latter, we observed that though before the seizure
the activity in one channel could be explained using a mixed embedding containing
also components from the antidiametrical channel, right after the seizure there were
no such components any more. Thus information flow can be coarsely detected from the
form of the embedding vector.

For a quantitative indication of information flow,
we developed the prediction measure $S$ and the information measure $R$, both making
use of the mixed and projected embedding vectors. These measures were able to detect different
degrees of coupling, and they were found to have good significance and power using
appropriate surrogate data testing.

Our embedding scheme is certainly not restricted
to bivariate time series and it is straightforward to apply it to more than two
time series in order
to investigate direct and indirect coupling among them.

\section*{Acknowledgments}
This paper is part of the 03ED748 research project, implemented within the framework of the ``Reinforcement Programme of Human Research Manpower''
(PENED) and co-financed at 90\% by National and Community Funds (25\% from the Greek Ministry of Development--General Secretariat of Research and
Technology and 75\% from E.U.--European Social Fund) and at 10\% by Rikshospitalet, Norway. We thank P\r{a}l G. Larsson for providing us with the EEG
data and for guiding us for the selection of EEG channels.

\end{document}